\documentclass[pra,aps,twocolumn,floatfix,nofootinbib,superscriptaddress]{revtex4-2}
\usepackage{amsmath}
\usepackage{amssymb}
\usepackage{amsthm}
\usepackage{bm}
\usepackage[colorlinks=true,
					  linkcolor=googleblue,
						citecolor=carmine,
						urlcolor=black]{hyperref}
\usepackage{float}
\usepackage{graphicx}
	\graphicspath{{figs/}}
\usepackage{overpic}
\usepackage{mathtools}
\usepackage{xcolor}
	\definecolor{carmine}{RGB}{150,0,24}
	\definecolor{googleblue}{RGB}{34,0,204}
\usepackage[caption=false]{subfig}
\usepackage{tikz}

\newcommand{\ket}[1]{\mbox{$ | #1 \rangle $}}

\newcommand{\crule}[3][black]{\textcolor{#1}{\rule{#2}{#3}}}
\newcommand*{\excluded}{\begin{tikzpicture}[scale=0.15]%
    \draw[fill opacity=1,fill=white,thick] (0,0) circle (0.7);
    \end{tikzpicture}}

\newtheorem*{theorem}{Observation}
\newtheorem*{result}{Result}

\begin{document}

\title{Proofs of network quantum nonlocality in continuous families of distributions}

\author{Alejandro Pozas-Kerstjens}
\affiliation{Instituto de Ciencias Matem\'aticas (CSIC-UAM-UC3M-UCM), 28049 Madrid, Spain}
\affiliation{Departamento de An\'alisis Matem\'atico, Universidad Complutense de Madrid, 28040 Madrid, Spain}
\author{Nicolas Gisin}
\affiliation{Group of Applied Physics, University of Geneva, 1211 Geneva 4, Switzerland}
\affiliation{Constructor University, Geneva, Switzerland}
\author{Marc-Olivier Renou}
\affiliation{ICFO - Institut de Ci\`encies Fot\`oniques, The Barcelona Institute of Science and Technology, 08860 Castelldefels (Barcelona), Spain}

\begin{abstract}
	The study of nonlocality in scenarios that depart from the bipartite Einstein-Podolsky-Rosen setup is allowing to uncover many fundamental features of quantum mechanics.
	Recently, an approach to building network-local models based on machine learning lead to the conjecture that the family of quantum triangle distributions of [Renou \textit{et al.}, \href{https://doi.org/10.1103/PhysRevLett.123.140401}{\color{googleblue}{Phys. Rev. Lett. \textbf{123}, 140401 (2019)}]} did not admit triangle-local models in a larger range than the original proof.
	We prove part of this conjecture in the affirmative.
	Our approach consists in reducing the family of original, four-outcome distributions to families of binary-outcome ones, and then using the inflation technique to prove that these families of binary-outcome distributions do not admit triangle-local models.
	This constitutes the first successful use of inflation in a proof of quantum nonlocality in networks whose nonlocality could not be proved with alternative methods.
	Moreover, we provide a method to extend proofs of network nonlocality in concrete distributions of a parametrized family to continuous ranges of the parameter.
	In the process, we produce a large collection of network Bell inequalities for the triangle scenario with binary outcomes, which are of independent interest.
\end{abstract}

\maketitle

One of the most striking consequences of Bell's theorem~\cite{BellTheorem} and its violation by quantum mechanics is that nature cannot be modeled using only local variables.
This phenomenon, known as quantum nonlocality, is now the basis of many protocols that process information encoded in quantum systems~\cite{cleve1997,May98,acin2007}.
In the last decade, both theoretical advances and experimental improvements have allowed to shift the focus to multipartite scenarios, called networks, where many independent sources distribute physical systems among different, overlapping collections of parties~\cite{Branciard2010,networkReview}.
The study of nonlocality in these network scenarios is generating a very exciting research field.
On one hand, it provides the framework needed to analyze the distribution of quantum information in complex networks (i.e., the quantum Internet~\cite{Kimble2008,Wehner2018,Kozlowski2019}).
On the other hand, the consideration of independent sources departs from the traditional Einstein-Podolsky-Rosen scenario, enabling a deeper analysis on the foundations of quantum mechanics~\cite{Gisin2020,renou2021complex,coiteuxroy2021PRL,coiteuxroy2021PRA,supic2022}.

\begin{figure}[h!]
	\flushleft
	\hspace{0.3cm}
	\begin{overpic}[width=0.48\columnwidth]{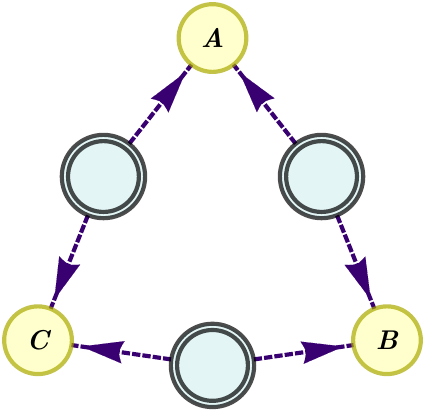}
		\put(44,9){\large$\psi^+$}
		\put(18,53){\large$\psi^+$}
		\put(70,53){\large$\psi^+$}
		\put(100,80){$P_u(a,b,c)$ is triangle-local}
		\put(140,70){$\Updownarrow$}
		\put(98,60){(i) $\exists\, q_u(i,j,k,t)$ compatible}
		\put(130,50){such that}
		\put(90,40){(ii) $q_u(i,j,k|t)$ are triangle-local}
		\put(15,-25){\crule{6.5cm}{1.5pt}}
		\put(14,-30){\crule{1.5pt}{0.5cm}}
		\put(8,-12){\large$\frac{1}{\sqrt{2}}$}
		\put(12,-27){\excluded}
		\put(37,-30){\crule{1.5pt}{0.5cm}}
		\put(25,-15){\large$0.7504$}
		\put(68,-30){\crule{1.5pt}{0.5cm}}
		\put(57,-15){\large$0.8101$}
		\put(108,-30){\crule{1.5pt}{0.5cm}}
		\put(98,-15){\large$0.8860$}
		\put(175,-27){\large$u$}
		\put(169,-30){\crule{1.5pt}{0.5cm}}
		\put(167.5,-15){\large$1$}
		\put(167,-27){\excluded}
		\put(108.5,-65){\crule{0.5pt}{1.5cm}}
		\put(108.5,-62){\vector(1,0){58}}
		\put(109.5,-62){\vector(-1,0){1}}
		\put(108.5,-62){\vector(-1,0){92}}
		\put(108,-62){\vector(1,0){1}}
		\put(114,-70){\small{Proved in \cite{renou2019genuine}}}
		\put(30,-70){\small{Conjectured in \cite{krivachy2019neural}}}
		\put(37.5,-52){\crule{0.5pt}{1.3cm}}
		\put(68.5,-52){\crule{0.5pt}{1.3cm}}
		\put(50,-50){\vector(-1,0){12}}
		\put(50,-50){\vector(1,0){18.5}}
		\put(36,-58){\small{This work}}
	\end{overpic}
	\vspace{3cm}
	\caption{(Top left) Realization of the family of quantum genuine triangle nonlocal distributions $P_u(a,b,c)$ of Ref.~\cite{renou2019genuine} [for completeness, also defined in Eq.~\eqref{eq:genuine}]. Each of the sources distributes a maximally entangled state $\ket{\psi^+}\,{\propto}\,\ket{01}\,{+}\,\ket{10}$, and each of the parties performs the four-outcome measurement given by $\{\ket{\bar{0}},\ket{\bar{1}_+},\ket{\bar{1}_-},\ket{\bar{2}}\}$ described in the main text. (Top right) Implications from the existence of a triangle-local model for $P_u(a,b,c)$, in the notation used in this work (see Appendix~\ref{app:equiv}). Reference \cite{renou2019genuine} used only condition (i) for asserting the nonlocality of $P_u(a,b,c)$. In this work we analyze when condition (ii) is not true. (Bottom) The current standing of the proofs of nonlocality for the family.
	}
	\label{fig:results}
\end{figure}
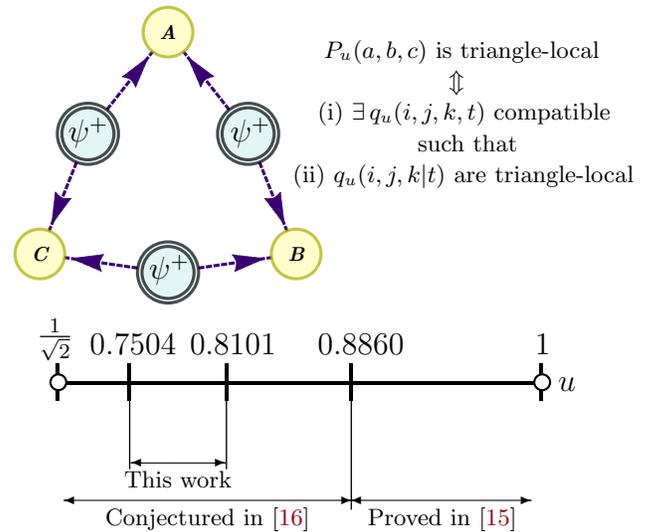

Nonlocality in networks is defined in a way analogous to the standard notion, namely by opposition to admitting a network-local model.
Network-local correlations are those that can be generated by assuming that the sources in the network distribute independent classical shared randomness, and the parties process the randomness they receive (potentially depending on a private choice of input) in order to generate an outcome.
Put as an example, consider the so-called triangle scenario, whereby three bipartite sources distribute systems to three separate parties forming a triangle-shaped structure (see Fig.~\ref{fig:results}).
Triangle-local correlations admit models of the form
\begin{align}
  P(a,b,c)=\int& \text{d}\alpha \text{d}\beta \text{d}\gamma\, \mu_{BC}(\alpha)\mu_{AC}(\beta)\mu_{AB}(\gamma) \notag\\
	&\times P_A(a|\beta,\gamma) P_B(b|\gamma,\alpha) P_C(c|\alpha,\beta),
  \label{eq:trianglelocal}
\end{align}
where $\alpha$, $\beta$ and $\gamma$ denote sources of classical shared randomness distributed with densities $\mu_{BC}(\alpha)$, $\mu_{AC}(\beta)$ and $\mu_{AB}(\gamma)$, respectively, and $P_A$, $P_B$ and $P_C$ represent local operations on the variables received by each party\footnote{Note that in Eq.~\eqref{eq:trianglelocal} the local operations do not depend on an input. These are, however, straightforward to insert if necessary.}.
Any distribution that cannot be generated by a suitable choice of $\mu_{AB}$, $\mu_{BC}$, $\mu_{AC}$, $P_A$, $P_B$ and $P_C$ is thus termed triangle-nonlocal.
For other networks, the definitions of network-local correlations and network nonlocality are analogous.

Many of the initial demonstrations of nonlocality in networks were based in standard bipartite Bell nonlocality.
For instance, Ref.~\cite{Gisin2017} showed that any entangled state that violates the CHSH inequality can be used to violate the bilocality inequality of Refs.~\cite{Branciard2010,Branciard2012}, and Ref.~\cite{Fritz2012} demonstrated that bipartite nonlocality can be ``disguised'' in networks, leading to network-nonlocal distributions.
Examples of this phenomenon motivated the search of stronger definitions for nonlocality in networks, that are in closer relation to the network structure~\cite{pozas2021fullnn} and give a more pivotal role to entangled measurements~\cite{supicGenuine}.

Recently, a family of quantum distributions in the triangle scenario [see Eq.~\eqref{eq:genuine} for its definition] was proven not to admit triangle-local models~\cite{renou2019genuine}.
The proof, based on token-counting arguments~\cite{Renou2020PRA,Renou2020PRL}, does not resemble any connection to methods employed in standard Bell nonlocality.
For this reason, it is commonly accepted as the first demonstration of genuinely quantum triangle nonlocality.
Unfortunately, the proof technique used in Ref.~\cite{renou2019genuine} is tailored to the particular family of distributions, and thus distributions that are close to but outside the family (such as those arising in experimental realizations~\cite{abiuso2021quantum}) cannot be proven to be triangle nonlocal using those methods.
This has motivated the search of numerical approaches to estimate noise tolerance.
Reference \cite{krivachy2019neural} uses a parametrization of triangle-local models based on machine learning, whereby the local responses of each of the parties are modeled as deep neural networks that process the systems received from the sources.
This enables the leverage of the deep learning toolbox to the search of local models in networks.
Surprisingly, this approach led the authors to conjecture that the family of distributions of Ref.~\cite{renou2019genuine} was triangle nonlocal well beyond the limits established by the original proof (see Fig.~\ref{fig:results}).
However, the inherent features of the formalism of deep learning prevent this conjecture from becoming a rigorous demonstration.

In this work, we prove the conjecture formulated in Ref.~\cite{krivachy2019neural} for a certain range of the parameter characterizing the family of distributions.
Our results are summarized in Fig.~\ref{fig:results}.
In a first step, we reduce the problem of compatibility of the four-outcome distribution family of Ref.~\cite{renou2019genuine} with models of the form of Eq.~\eqref{eq:trianglelocal} to the compatibility of families of binary-outcome distributions.
Later, we use the inflation technique~\cite{wolfe2019inflation} to rule out any such model for the binary-outcome families, which imply that the four-outcome distributions do not admit triangle-local models of the form of Eq.~\eqref{eq:trianglelocal}.
To our knowledge, this work constitutes the first successful use of inflation in a proof of quantum nonlocality in networks for distributions whose nonlocality could not be proved with alternative methods (e.g., those in Refs.~\cite{Chaves2015,Weilenmann2017,Pozas2019}).

In the process, we produce a large family of network Bell inequalities for the binary-outcome triangle network.
This family has the potential of deciding whether the set of binary-outcome quantum distributions in the triangle can be completely characterized in terms of triangle-local models, which is an important conjecture in the field~\cite{Fraser2018}.
However, due to the outcome-reduction process of our proof being tailored to the family of Ref.~\cite{renou2019genuine}, the question of noise resistance remains open.

\paragraph*{The distributions.---}
The family of distributions analyzed in Ref.~\cite{renou2019genuine} can be generated via measurements on bipartite quantum states distributed according to the triangle scenario, and it has a simple interpretation in terms of excitation counts~\cite{Renou2020PRA,Renou2020PRL,abiuso2021quantum}.
Consider the three sources in the triangle scenario distributing Bell states \mbox{$\ket{\psi^+}\,{=}\,(\ket{01}\,{+}\,\ket{10})/\sqrt{2}$}.
These states represent an excitation being sent to the left ($\ket{10}$) or to the right ($\ket{01}$).
Then, each of the parties measures the number of excitations received.
Receiving zero or two excitations is associated to the states $\ket{\bar{0}}\,{=}\,\ket{00}$ and $\ket{\bar{2}}\,{=}\,\ket{11}$, and the space spanned by the one-excitation events is two-dimensional.
The parties measure one-excitation events in superposition, determined by the projectors associated to the states $\ket{\bar{1}_+}\,{=}\,u\ket{01}\,{+}\,v\ket{10}$ and $\ket{\bar{1}_-}\,{=}\,v\ket{01}\,{-}\,u\ket{10}$, with $1/\sqrt{2}\,{\leq}\, u \,{\leq}\, 1$ and $v\,{=}\,\sqrt{1-u^2}$.
The result of all the parties performing this four-outcome measurement on the shares received is a family of distributions, $P_u(a,b,c)$, given by
\begin{equation}
  \begin{split}
    P_u(\bar{0},\bar{1}_-,\bar{2})=P_u(\bar{0},\bar{2},\bar{1}_+)&=\frac{u^2}{8}, \\
    P_u(\bar{0},\bar{1}_+,\bar{2})=P_u(\bar{0},\bar{2},\bar{1}_-)&=\frac{v^2}{8}, \\
    P_u(\bar{1}_j,\bar{1}_j,\bar{1}_j)&=\frac{[v^3+j\,u^3]^2}{8}, \\
    P_u(\bar{1}_j,\bar{1}_j,\bar{1}_k)&=\frac{u^2v^2}{8}[u+k\,v]^2,
  \end{split}
  \label{eq:genuine}
\end{equation}
plus cyclic permutations, and all the remaining probabilities being $0$.

Reference \cite{renou2019genuine} proved that, for every value of $u$, the three-outcome distribution obtained when every party coarse grains their respective outcomes $\bar{1}_i$, $\bar{1}_j$ and $\bar{1}_k$, into a single output, $\bar{1}$, admits an essentially unique triangle-local model with binary, uniformly random hidden variables.
Then, when the full $P_u(a,b,c)$ admits a triangle-local model, this must be compatible with the former, and in particular with a new variable $t$ containing information about the random hidden variables used in the coarse-grained triangle-local model.
This allows one to introduce a distribution $q_u(i,j,k,t)$, with binary outcomes and related to $P_u(a,b,c)$ via linear constraints, whose existence is equivalent to $P_u(a,b,c)$ satisfying Eq.~\eqref{eq:trianglelocal}.
We prove this statement in appendices \ref{app:polytope} and \ref{app:equiv}.

Some of the marginals of this distribution are fixed by $P_u(a,b,c)$, leading to the form (derived explicitly in Appendix~\ref{app:polytope})
\begin{align}
  q_u(i,j,k,t) = \frac{1}{2^4}&\left[1 + (u^2-v^2)^2(ij+jk+ki) +\right. \notag\\
  & + (u^2-v^2)(i+j+k)t + 8u^3v^3ijk \notag\\
  & + t(ij F_{AB}+jk F_{BC}+ki F_{AC}) \notag\\
	& + \left.ijkt F_{ABC}\right].
	\label{eq:q}
\end{align}
Note that the expression contains four free parameters, $F_{AB},F_{BC},F_{AC},F_{ABC}\,{\in}\,[-1,1]$.
The requirement that all probabilities in Eq.~\eqref{eq:q} are positive is the condition that Ref.~\cite{renou2019genuine} uses to find that a triangle-local model for $P_u(a,b,c)$ does not exist for $u\geq \sqrt{\frac{-3 + (9 + 6 \sqrt{3})^{2/3}}{2 (9 + 6 \sqrt{3})^{1/3}}}\equiv u_0\approx 0.8860$ (see Fig.~\ref{fig:results} and Appendix~\ref{app:polytope}).
Indeed, for $u_0<u<1$, at least one of the probabilities of Eq.~\eqref{eq:q} is negative for any value of $F_{AB},F_{BC},F_{AC},F_{ABC}\,{\in}\,[-1,1]$.

In contrast, for $1/\sqrt{2}\,{<}\,u\,{<}\,u_0$, there always exist values of $F_{AB}$, $F_{BC}$, $F_{AC}$ and $F_{ABC}$ that make all probabilities in Eq.~\eqref{eq:q} positive.
This does not mean, however, that the corresponding $P_u(a,b,c)$ admits a model of the form of Eq.~\eqref{eq:trianglelocal}.
In Appendix~\ref{app:equiv} we show that this is the case if and only if there exist specific values for $F_{AB}$, $F_{BC}$, $F_{AC}$ and $F_{ABC}$ such that $q_u(i,j,k,t)$  is a proper probability distribution and the two conditionals $q_u^t(i,j,k)\,{\coloneqq}\,q_u(i,j,k|t)$ admit triangle-local models of the form of Eq.~\eqref{eq:trianglelocal} as well.
Reversing the argument, finding that either $q_u^+(i,j,k)$ or $q_u^-(i,j,k)$ is triangle nonlocal for all allowed values of $F_{AB}$, $F_{BC}$, $F_{AC}$, and $F_{ABC}$ is proof that $P_u(a,b,c)$ is triangle nonlocal.
In the following, thus, we find a range of $u$ where $q_u^+(i,j,k)$ and $q_u^-(i,j,k)$ do not admit triangle-local models.

\paragraph*{Proof for discrete values $u\,{<}\,u_0$.---}
We now show that, for $0.7504 \,{\leq}\, u \,{\leq}\, 0.8101$, at least one of the $q^t_u(i,j,k)$ for $t\,{=}\,\pm1$ does not admit such a triangle-local model in the whole region where $q_u(i,j,k,t)$ is well defined.
This region is described by the positivity of Eq.~\eqref{eq:q} for every tuple $(i,j,k,t)$.
When the value of $u$ is fixed, the region is a polytope in $\mathbb{R}^4$, which we characterize explicitly in Appendix~\ref{app:polytope}.
Let us denote this polytope $\mathcal{P}_u$.

For $t\,{=}\,+1$, and at for $1/\sqrt{2}\,{\leq}\, u \,{\lesssim}\, 0.8457$, there always exists a distribution in $\mathcal{P}_u$ that admits a triangle-local model.
We describe this distribution and its triangle-local model in Appendix \ref{app:qplusmodel}.
This means that, at least for that range, the only way to prove the nonlocality of $P_u(a,b,c)$ is proving that $q^-(i,j,k)$ is triangle nonlocal.
We do so via the inflation technique for causal compatibility~\cite{wolfe2019inflation}.

Inflation provides a hierarchy of necessary conditions, characterized by linear programming problems, that are satisfied by correlations $q$ admitting network-local models~\cite{navascues2017convergence}.
Briefly, inflation is a technique by which the sources and measurement devices of a network are copied several times and arranged in different configurations.
By studying the correlations $p_\text{inf}$ that arise in these configurations, one can find necessary conditions for correlations $q$ to be generated in the original network.
Note, however, that inflation is only a theoretical tool and does not mean that the actual network of interest is changed.
The fact that, for a particular distribution under study, the necessary conditions are not satisfied at some level of the inflation hierarchy constitutes a rigorous proof that such distribution does not admit a network local model.

In this work we consider the compatibility of $q_u^-(i,j,k)$ with triangle-local models through the so-called web inflation of the triangle network (see Fig.~\ref{fig:WebInflation} in Appendix~\ref{app:inflation}), where distributions compatible with the inflation are related to $q_u^-(i,j,k)$ via three types of constraints.
Firstly, we impose the \textit{hierarchy} constraints (which are sufficient to prove the asymptotic convergence of inflation \cite{navascues2017convergence}).
These linear conditions relate the marginals of the inflation distribution $p_\text{inf}$ over subnetworks which completely reproduce the original network to the distribution $q_u^-(i,j,k)$.
Secondly, we impose the so-called \textit{higher-degree relations} of Ref.~\cite{navascues2017convergence} (also used in earlier works; see for instance the discussion in \cite[Example 2]{wolfe2019inflation}).
These linear constraints relate marginals of $p_\text{inf}$ to polynomials in the original distribution $q_u^-(i,j,k)$ of degree higher than the inflation level.
Finally, we also impose \textit{linearized polynomial identification} (LPI) constraints.
These are, \textit{a priori}, nonlinear polynomial relations between the elements of the probability distribution in the inflation, which cannot be enforced in linear programming.
However, once the remaining constraints have been imposed for a given distribution, some of the factors are substituted by known probabilities, rendering the constraints linear in $p_\text{inf}$.
LPI constraints have been thoroughly used in the literature \cite{alexThesis,Gisin2020,Boreiri2022} since they appear to constrain significantly the sets of compatible correlations.
In Appendix~\ref{app:inflation} we provide further information on the inflation employed and the different types of linear constraints that we use, with concrete examples.

For a given $u$, the compatibility of $q^-_u(i,j,k)$ with triangle-local models is thus relaxed to a linear program $[\text{find }\bm{x}$ $\text{ such that }A\cdot\bm{x}\geq\bm{b}]$, where the unknown $\bm{x}\equiv p_\text{inf}$ contains the probabilities of events in the inflation scenario, $A$ is the matrix of coefficients of the linear inequalities that the probabilities are subject to (some depending on $q^-_u(i,j,k)$ due to the LPI constraints), and $\bm{b}$ contains the information about $q^-_u(i,j,k)$.

The nonexistence of a solution to a linear program is associated to a witness, as stated by Farkas' lemma~\cite{FarkasLemma}.
Namely, if the linear program $A\cdot\bm{x}\geq\bm{b}$ does not have any solution $\bm{x}$, then there exists a vector $\bm{y}\geq 0$ that satisfies $\bm{y}\cdot A =\bm{0}$ and $\bm{y}\cdot\bm{b} > 0$, thus witnessing the infeasibility of the program.
Since in traditional problems in nonlocality the matrix $A$ is constant, the expression $\bm{y}\cdot\bm{b} > 0$ can be understood as a nonlocality witness.
In fact, this approach is standard for finding Bell inequalities from results on the inexistence of local models~\cite{baccari2017detection,Fraser2018,pozas2021fullnn}.
For the inflation used (given in Fig.~\ref{fig:WebInflation} in Appendix~\ref{app:inflation}) and the distributions $q_u^-(i,j,k)$ created from conditioning Eq.~\eqref{eq:q}, the matrix $A$ depends on $u$ but not on $F_{AB}$, $F_{AC}$, $F_{BC}$, or $F_{ABC}$, and hence we will denote it as $A_u$.
Therefore, Farkas' lemma provides a witness $\bm{y}_u$ valid for fixed $u$, because only for a given $u$ is the condition $\bm{y}\cdot A_u =\bm{0}$ satisfied.

By discretizing the range of $u$ we find, for every point $0.7504\,{\leq}\,u\,{\leq}\,0.8101$ in the discretization, an infeasibility certificate $\bm{y}_u$ that identifies every distribution in the corresponding $\mathcal{P}_u$ as incompatible with a triangle-local model (an illustration of one such certificate appears in Fig.~\ref{fig:constraintCertificates} in Appendix~\ref{app:inflation}).
This certificate is obtained when analyzing the compatibility of the distribution $q^-_u(i,j,k)$ derived from Eq.~\eqref{eq:q} for $F_{AB}\,{=}\,F_{BC}\,{=}\,F_{AC}\,{=}2v^2(u^2-v^2)$, $F_{ABC}\,{=}\,1-8u^3v^3$, which is a vertex of $\mathcal{P}_u$ for $1/\sqrt{2}\,{\leq}\, u \,{\lesssim}\, 0.8457$.
In the computational appendix~\cite{compAppendix} we provide computer codes that find such witnesses by solving the respective linear programs.
Each of these witnesses identifies that, for its corresponding value of $u$, all distributions in the family $q^-_u(i,j,k)$ are triangle nonlocal.
By the arguments above, this implies that $P_u(a,b,c)$ is triangle-nonlocal for the corresponding values of $u$.

Outside the range, it is still possible to find values of $F_{AB}$, $F_{BC}$, $F_{AC}$, and $F_{ABC}$ for which $q^-_u(i,j,k)$ is triangle nonlocal.
The associated witnesses, however, do not identify nonlocality for all values of $F_{AB}$, $F_{BC}$, $F_{AC}$ and $F_{ABC}$, and so they do not guarantee that $P_u(a,b,c)$ is triangle nonlocal.
Since inflation provides outer approximations of the sets of compatible correlations, and $\mathcal{P}_u$ does not qualitatively change at the boundaries of the interval, it is possible that the cause of this behavior is of a purely computational nature.
Thus, increasing the computational capabilities in order to consider larger inflations is likely to push the boundaries of the interval where nonlocality can be proven.

\paragraph*{Extending to the continuum.---}
The fact that we can identify that $q^-_u(i,j,k)$ does not admit a triangle-local model for discrete values of $0.7504\,{\leq}\,u\,{\leq}\,0.8101$ does not rule out the possibility of the existence of discrete points or small regions within that interval where a model exists.
This is, in fact, a common problem that appears when using numerical techniques to characterize nonlocality.
To address this issue, we make use of the following observation, which is a consequence of a more general statement that stems from Farkas' lemma and that we discuss in Appendix~\ref{app:continuum}:
\begin{theorem}[Noise-tolerant infeas. certificate]
		For the matrices $A$ and the vectors $\bm{b}$ that characterize the linear programs described by Eq.~\eqref{LP}, and for the distributions that arise from conditioning Eq.~\eqref{eq:q} over $t$, if $\bm{y}_u$ is the certificate of infeasibility at some fixed $u$, the inequality $\bm{y}_u\cdot \bm{b}_{u'} > \max(\bm{y}_u\cdot A_{u'})$ is a witness of triangle nonlocality for a continuous range of $u'$ centered around $u$.
		\label{continuous}
\end{theorem}

It must be noted that the observation above is a consequence of a more general statement, namely that using the inequality $\bm{y}\cdot \bm{b} > \max(\bm{y}\cdot A)$ as a certificate of infeasibility is possible for any linear program with $\|\bm{x}\|\leq B$ if the right-hand side is multiplied by $B$.
Using this, in the computational appendix~\cite{compAppendix} we provide a computer program that, recursively, proves the triangle nonlocality of $P_{u_\text{init}}(a,b,c)$ for a particular value of $u_\text{init}$, computes the extremum $u_\text{fin}$ for which the witness $\bm{y}_{u_\text{init}}$ identifies that $P_u(a,b,c)$ is triangle nonlocal in the entire interval $[u_\text{init},u_\text{fin}]$, and sets $u_\text{init}\,{=}\,u_\text{fin}$ to begin again.
Starting at $u=0.7504$, the program halts at $0.8101$, producing along the way a collection of witnesses that excludes the whole interval $[0.7504, 0.8101]$.
The program produces a total of 892 witnesses in the whole interval (all of them stored in \cite{compAppendix}), each of which is able to identify the triangle nonlocality of $P_u(a,b,c)$ for a range of $u$ of width $1.3\times10^{-4}$ in average.
This collection of certificates proves that
\begin{result}
	$P_u(a,b,c)$ is triangle nonlocal in the whole interval $0.7504\,{\leq}\, u \,{\leq}\, 0.8101$.
\end{result}

\paragraph*{Discussion.---}
In this work we rigorously prove a conjecture initially formulated by applying machine learning techniques to network nonlocality, namely, that the family of quantum triangle distributions of Ref.~\cite{renou2019genuine} is also triangle nonlocal below $u_0\,{\approx}\, 0.8860$ calculated in the original proof.
The current standing, captured in Fig.~\ref{fig:results}, is that the family of distributions is triangle nonlocal for $u\,{\in}\,[0.7504,0.8101]\cup]u_0,1)$.
For the remaining region, namely $u\,{\in}\,]1/\sqrt{2},0.7504)\cup(0.8101,u_0[$, the question remains open.
The intuitions developed via the machine learning approach of Ref.~\cite{krivachy2019neural} point to a positive answer for the whole range $u\,{\in}\,]1/\sqrt{2},u_0[$.
Because the characterization of the set of triangle-local distributions offered by inflation is an outer approximation of the real one, and the polytope $\mathcal{P}_u$ used in our proof does not qualitatively change in the endpoints found, we expect that the conjecture is true for the full range and its proof can be asymptotically reached with increased computational capabilities in order to work with larger inflations.

Beyond the family of distributions $P_u(a,b,c)$, it was proven in Ref.~\cite{renou2019genuine} (Fig.~\ref{fig:polytope}) that the method given in this work can be applied to token-counting distributions generated from partially entangled states.
This method was also extended to a large class of networks beyond the triangle scenario in Refs.~\cite{Renou2020PRA,Renou2020PRL}.
The new method developed in this work can be used to obtain new nonlocal distributions for these configurations, extending the range of parameters for which nonlocality can be proven.
Moreover, the proof technique used here ``recycles'' infeasibility certificates of a given linear program to witness infeasibility of others.
This technique goes beyond nonlocality and can be used wherever linear programming is relevant.

Our construction also provides a large family of polynomial, network Bell inequalities for binary-outcome distributions in the triangle network, which we present in Ref.~\cite{compAppendix}.
Despite initially being only applicable for the family of binary-outcome distributions under study, we have extended them to be valid for arbitrary binary-outcome distributions by using the discussion in Appendix~\ref{app:continuum}.
Extending the range of applicability of Bell inequalities obtained from infeasibility certificates is a novel approach, which can help in making rigorous statements from numerical results when addressing nonlocality in arbitrary networks, or in problems beyond network nonlocality where linear programming becomes relevant.
However, because the reduction from four-outcome to binary-outcome distributions is exclusive to the family of distributions in Ref.~\cite{renou2019genuine}, these inequalities cannot be directly understood as noise-robust certificates of triangle nonlocality for the family $P_u(a,b,c)$.

More generally, it is not known whether the derived inequalities can be violated in quantum mechanics.
Finding such a violation in any of them would decide in the positive another long-standing open question in the field, namely whether the binary-outcome triangle scenario without inputs supports network nonlocality.
In particular, it is interesting to test the distributions generated out of maximally network-nonlocal boxes of Ref.~\cite{Bancal2021}.
Moreover, a very interesting question that we now open is the existence of a family of network Bell inequalities, with the coefficients being continuous functions of $u$, that identify the corresponding distribution as triangle nonlocal.
This family could provide insights into noise-robust witnesses of the nonlocality of the family of Ref.~\cite{renou2019genuine}.
A potentially promising starting point is the distribution for $u\,{=}\,55/73\,{\approx}\,0.7534$.
Since $(48,55,73)$ is a Pythagorean triple, the associated inflation problem \eqref{LP} and solution are formulated exactly in terms of rational numbers.

In a broader view, our results can be framed in the context of \textit{artificial intelligence augmentation}~\cite{carter2017AIA}, which understands that the ultimate use of machine learning in science will not be replacing humans in solving complicated problems, but becoming a tool for understanding where interesting problems lie.
A paradigmatic example of this approach to physics problems is the automated analysis of phase diagrams, which has suggested the existence of new phases in various models~\cite{venderley2018,kottmann2020}.
In this picture, the machine learning approach of Ref.~\cite{krivachy2019neural} pointed to an interesting question, which in turn produced results that have increased our understanding of triangle nonlocality, and provided with new network Bell inequalities that can help us in advancing other problems in the field.

\paragraph*{Acknowledgments.---}	A.~P.-K.~is supported by the European Union's Horizon 2020 research and innovation programme-grant agreement No. 648913, the Spanish Ministry of Science and Innovation through the ``Severo Ochoa Programme for Centres of Excellence in R\&D'' (CEX2019-000904-S) and PID2020-113523GB-I00, and Comunidad de Madrid through QUITEMAD-CM P2018/TCS-4342.
N.~G.~is supported by the Swiss National Science Foundation via the National Centres of Competence in Research (NCCR)-SwissMap.
M.-O.~R.~is supported by the Swiss National Fund Early Mobility Grant P2GEP2\_191444.

\bibliographystyle{apsrev4-2}
\bibliography{references}

\begin{thebibliography}{39}%
\makeatletter
\providecommand \@ifxundefined [1]{%
 \@ifx{#1\undefined}
}%
\providecommand \@ifnum [1]{%
 \ifnum #1\expandafter \@firstoftwo
 \else \expandafter \@secondoftwo
 \fi
}%
\providecommand \@ifx [1]{%
 \ifx #1\expandafter \@firstoftwo
 \else \expandafter \@secondoftwo
 \fi
}%
\providecommand \natexlab [1]{#1}%
\providecommand \enquote  [1]{``#1''}%
\providecommand \bibnamefont  [1]{#1}%
\providecommand \bibfnamefont [1]{#1}%
\providecommand \citenamefont [1]{#1}%
\providecommand \href@noop [0]{\@secondoftwo}%
\providecommand \href [0]{\begingroup \@sanitize@url \@href}%
\providecommand \@href[1]{\@@startlink{#1}\@@href}%
\providecommand \@@href[1]{\endgroup#1\@@endlink}%
\providecommand \@sanitize@url [0]{\catcode `\\12\catcode `\$12\catcode
  `\&12\catcode `\#12\catcode `\^12\catcode `\_12\catcode `\%12\relax}%
\providecommand \@@startlink[1]{}%
\providecommand \@@endlink[0]{}%
\providecommand \url  [0]{\begingroup\@sanitize@url \@url }%
\providecommand \@url [1]{\endgroup\@href {#1}{\urlprefix }}%
\providecommand \urlprefix  [0]{URL }%
\providecommand \Eprint [0]{\href }%
\providecommand \doibase [0]{https://doi.org/}%
\providecommand \selectlanguage [0]{\@gobble}%
\providecommand \bibinfo  [0]{\@secondoftwo}%
\providecommand \bibfield  [0]{\@secondoftwo}%
\providecommand \translation [1]{[#1]}%
\providecommand \BibitemOpen [0]{}%
\providecommand \bibitemStop [0]{}%
\providecommand \bibitemNoStop [0]{.\EOS\space}%
\providecommand \EOS [0]{\spacefactor3000\relax}%
\providecommand \BibitemShut  [1]{\csname bibitem#1\endcsname}%
\let\auto@bib@innerbib\@empty
\bibitem [{\citenamefont {Bell}(1964)}]{BellTheorem}%
  \BibitemOpen
  \bibfield  {author} {\bibinfo {author} {\bibfnamefont {J.~S.}\ \bibnamefont
  {Bell}},\ }\href {https://doi.org/10.1103/PhysicsPhysiqueFizika.1.195}
  {\bibfield  {journal} {\bibinfo  {journal} {Physics Physique Fizika}\
  }\textbf {\bibinfo {volume} {1}},\ \bibinfo {pages} {195} (\bibinfo {year}
  {1964})}\BibitemShut {NoStop}%
\bibitem [{\citenamefont {Cleve}\ and\ \citenamefont
  {Buhrman}(1997)}]{cleve1997}%
  \BibitemOpen
  \bibfield  {author} {\bibinfo {author} {\bibfnamefont {R.}~\bibnamefont
  {Cleve}}\ and\ \bibinfo {author} {\bibfnamefont {H.}~\bibnamefont
  {Buhrman}},\ }\href {https://doi.org/10.1103/PhysRevA.56.1201} {\bibfield
  {journal} {\bibinfo  {journal} {Phys. Rev. A}\ }\textbf {\bibinfo {volume}
  {56}},\ \bibinfo {pages} {1201} (\bibinfo {year} {1997})}\BibitemShut
  {NoStop}%
\bibitem [{\citenamefont {Mayers}\ and\ \citenamefont {Yao}(1998)}]{May98}%
  \BibitemOpen
  \bibfield  {author} {\bibinfo {author} {\bibfnamefont {D.}~\bibnamefont
  {Mayers}}\ and\ \bibinfo {author} {\bibfnamefont {A.}~\bibnamefont {Yao}},\
  }in\ \href {https://doi.org/10.1109/SFCS.1998.743501} {\emph {\bibinfo
  {booktitle} {Proceedings 39th Annual Symposium on Foundations of Computer
  Science (FOCS)}}}\ (\bibinfo {year} {1998})\ pp.\ \bibinfo {pages}
  {503--509}\BibitemShut {NoStop}%
\bibitem [{\citenamefont {Ac\'{\i}n}\ \emph {et~al.}(2007)\citenamefont
  {Ac\'{\i}n}, \citenamefont {Brunner}, \citenamefont {Gisin}, \citenamefont
  {Massar}, \citenamefont {Pironio},\ and\ \citenamefont {Scarani}}]{acin2007}%
  \BibitemOpen
  \bibfield  {author} {\bibinfo {author} {\bibfnamefont {A.}~\bibnamefont
  {Ac\'{\i}n}}, \bibinfo {author} {\bibfnamefont {N.}~\bibnamefont {Brunner}},
  \bibinfo {author} {\bibfnamefont {N.}~\bibnamefont {Gisin}}, \bibinfo
  {author} {\bibfnamefont {S.}~\bibnamefont {Massar}}, \bibinfo {author}
  {\bibfnamefont {S.}~\bibnamefont {Pironio}},\ and\ \bibinfo {author}
  {\bibfnamefont {V.}~\bibnamefont {Scarani}},\ }\href
  {https://doi.org/10.1103/PhysRevLett.98.230501} {\bibfield  {journal}
  {\bibinfo  {journal} {Phys. Rev. Lett.}\ }\textbf {\bibinfo {volume} {98}},\
  \bibinfo {pages} {230501} (\bibinfo {year} {2007})}\BibitemShut {NoStop}%
\bibitem [{\citenamefont {Branciard}\ \emph {et~al.}(2010)\citenamefont
  {Branciard}, \citenamefont {Gisin},\ and\ \citenamefont
  {Pironio}}]{Branciard2010}%
  \BibitemOpen
  \bibfield  {author} {\bibinfo {author} {\bibfnamefont {C.}~\bibnamefont
  {Branciard}}, \bibinfo {author} {\bibfnamefont {N.}~\bibnamefont {Gisin}},\
  and\ \bibinfo {author} {\bibfnamefont {S.}~\bibnamefont {Pironio}},\ }\href
  {https://doi.org/10.1103/PhysRevLett.104.170401} {\bibfield  {journal}
  {\bibinfo  {journal} {Phys. Rev. Lett.}\ }\textbf {\bibinfo {volume} {104}},\
  \bibinfo {pages} {170401} (\bibinfo {year} {2010})}\BibitemShut {NoStop}%
\bibitem [{\citenamefont {Tavakoli}\ \emph {et~al.}(2022)\citenamefont
  {Tavakoli}, \citenamefont {Pozas-Kerstjens}, \citenamefont {Luo},\ and\
  \citenamefont {Renou}}]{networkReview}%
  \BibitemOpen
  \bibfield  {author} {\bibinfo {author} {\bibfnamefont {A.}~\bibnamefont
  {Tavakoli}}, \bibinfo {author} {\bibfnamefont {A.}~\bibnamefont
  {Pozas-Kerstjens}}, \bibinfo {author} {\bibfnamefont {M.-X.}\ \bibnamefont
  {Luo}},\ and\ \bibinfo {author} {\bibfnamefont {M.-O.}\ \bibnamefont
  {Renou}},\ }\href {https://doi.org/10.1088/1361-6633/ac41bb} {\bibfield
  {journal} {\bibinfo  {journal} {Rep. Prog. Phys.}\ }\textbf {\bibinfo
  {volume} {85}},\ \bibinfo {pages} {056001} (\bibinfo {year}
  {2022})}\BibitemShut {NoStop}%
\bibitem [{\citenamefont {Kimble}(2008)}]{Kimble2008}%
  \BibitemOpen
  \bibfield  {author} {\bibinfo {author} {\bibfnamefont {H.~J.}\ \bibnamefont
  {Kimble}},\ }\href {https://doi.org/10.1038/nature07127} {\bibfield
  {journal} {\bibinfo  {journal} {Nature}\ }\textbf {\bibinfo {volume} {453}},\
  \bibinfo {pages} {1023} (\bibinfo {year} {2008})}\BibitemShut {NoStop}%
\bibitem [{\citenamefont {Wehner}\ \emph {et~al.}(2018)\citenamefont {Wehner},
  \citenamefont {Elkouss},\ and\ \citenamefont {Hanson}}]{Wehner2018}%
  \BibitemOpen
  \bibfield  {author} {\bibinfo {author} {\bibfnamefont {S.}~\bibnamefont
  {Wehner}}, \bibinfo {author} {\bibfnamefont {D.}~\bibnamefont {Elkouss}},\
  and\ \bibinfo {author} {\bibfnamefont {R.}~\bibnamefont {Hanson}},\ }\href
  {https://science.sciencemag.org/content/362/6412/eaam9288} {\bibfield
  {journal} {\bibinfo  {journal} {Science}\ }\textbf {\bibinfo {volume}
  {362}},\ \bibinfo {pages} {6412} (\bibinfo {year} {2018})}\BibitemShut
  {NoStop}%
\bibitem [{\citenamefont {Kozlowski}\ and\ \citenamefont
  {Wehner}(2019)}]{Kozlowski2019}%
  \BibitemOpen
  \bibfield  {author} {\bibinfo {author} {\bibfnamefont {W.}~\bibnamefont
  {Kozlowski}}\ and\ \bibinfo {author} {\bibfnamefont {S.}~\bibnamefont
  {Wehner}},\ }in\ \href {https://doi.org/10.1145/3345312.3345497} {\emph
  {\bibinfo {booktitle} {Proceedings of the Sixth Annual ACM International
  Conference on Nanoscale Computing and Communication}}},\ \bibinfo {series and
  number} {NANOCOM '19}\ (\bibinfo  {publisher} {Association for Computing
  Machinery},\ \bibinfo {address} {New York, NY, USA},\ \bibinfo {year}
  {2019})\BibitemShut {NoStop}%
\bibitem [{\citenamefont {Gisin}\ \emph {et~al.}(2020)\citenamefont {Gisin},
  \citenamefont {Bancal}, \citenamefont {Cai}, \citenamefont {Remy},
  \citenamefont {Tavakoli}, \citenamefont {Zambrini~Cruzeiro}, \citenamefont
  {Popescu},\ and\ \citenamefont {Brunner}}]{Gisin2020}%
  \BibitemOpen
  \bibfield  {author} {\bibinfo {author} {\bibfnamefont {N.}~\bibnamefont
  {Gisin}}, \bibinfo {author} {\bibfnamefont {J.-D.}\ \bibnamefont {Bancal}},
  \bibinfo {author} {\bibfnamefont {Y.}~\bibnamefont {Cai}}, \bibinfo {author}
  {\bibfnamefont {P.}~\bibnamefont {Remy}}, \bibinfo {author} {\bibfnamefont
  {A.}~\bibnamefont {Tavakoli}}, \bibinfo {author} {\bibfnamefont
  {E.}~\bibnamefont {Zambrini~Cruzeiro}}, \bibinfo {author} {\bibfnamefont
  {S.}~\bibnamefont {Popescu}},\ and\ \bibinfo {author} {\bibfnamefont
  {N.}~\bibnamefont {Brunner}},\ }\href
  {https://doi.org/10.1038/s41467-020-16137-4} {\bibfield  {journal} {\bibinfo
  {journal} {Nat. Commun.}\ }\textbf {\bibinfo {volume} {11}},\ \bibinfo
  {pages} {2378} (\bibinfo {year} {2020})}\BibitemShut {NoStop}%
\bibitem [{\citenamefont {Renou}\ \emph {et~al.}(2021)\citenamefont {Renou},
  \citenamefont {Trillo}, \citenamefont {Weilenmann}, \citenamefont {Thinh},
  \citenamefont {Tavakoli}, \citenamefont {Gisin}, \citenamefont {Acin},\ and\
  \citenamefont {Navascu\'es}}]{renou2021complex}%
  \BibitemOpen
  \bibfield  {author} {\bibinfo {author} {\bibfnamefont {M.-O.}\ \bibnamefont
  {Renou}}, \bibinfo {author} {\bibfnamefont {D.}~\bibnamefont {Trillo}},
  \bibinfo {author} {\bibfnamefont {M.}~\bibnamefont {Weilenmann}}, \bibinfo
  {author} {\bibfnamefont {L.~P.}\ \bibnamefont {Thinh}}, \bibinfo {author}
  {\bibfnamefont {A.}~\bibnamefont {Tavakoli}}, \bibinfo {author}
  {\bibfnamefont {N.}~\bibnamefont {Gisin}}, \bibinfo {author} {\bibfnamefont
  {A.}~\bibnamefont {Acin}},\ and\ \bibinfo {author} {\bibfnamefont
  {M.}~\bibnamefont {Navascu\'es}},\ }\href
  {https://doi.org/10.1038/s41586-021-04160-4} {\bibfield  {journal} {\bibinfo
  {journal} {Nature}\ }\textbf {\bibinfo {volume} {600}},\ \bibinfo {pages}
  {625} (\bibinfo {year} {2021})}\BibitemShut {NoStop}%
\bibitem [{\citenamefont {Coiteux-Roy}\ \emph
  {et~al.}(2021{\natexlab{a}})\citenamefont {Coiteux-Roy}, \citenamefont
  {Wolfe},\ and\ \citenamefont {Renou}}]{coiteuxroy2021PRL}%
  \BibitemOpen
  \bibfield  {author} {\bibinfo {author} {\bibfnamefont {X.}~\bibnamefont
  {Coiteux-Roy}}, \bibinfo {author} {\bibfnamefont {E.}~\bibnamefont {Wolfe}},\
  and\ \bibinfo {author} {\bibfnamefont {M.-O.}\ \bibnamefont {Renou}},\ }\href
  {https://doi.org/10.1103/PhysRevLett.127.200401} {\bibfield  {journal}
  {\bibinfo  {journal} {Phys. Rev. Lett.}\ }\textbf {\bibinfo {volume} {127}},\
  \bibinfo {pages} {200401} (\bibinfo {year} {2021}{\natexlab{a}})}\BibitemShut
  {NoStop}%
\bibitem [{\citenamefont {Coiteux-Roy}\ \emph
  {et~al.}(2021{\natexlab{b}})\citenamefont {Coiteux-Roy}, \citenamefont
  {Wolfe},\ and\ \citenamefont {Renou}}]{coiteuxroy2021PRA}%
  \BibitemOpen
  \bibfield  {author} {\bibinfo {author} {\bibfnamefont {X.}~\bibnamefont
  {Coiteux-Roy}}, \bibinfo {author} {\bibfnamefont {E.}~\bibnamefont {Wolfe}},\
  and\ \bibinfo {author} {\bibfnamefont {M.-O.}\ \bibnamefont {Renou}},\ }\href
  {https://doi.org/10.1103/PhysRevA.104.052207} {\bibfield  {journal} {\bibinfo
   {journal} {Phys. Rev. A}\ }\textbf {\bibinfo {volume} {104}},\ \bibinfo
  {pages} {052207} (\bibinfo {year} {2021}{\natexlab{b}})}\BibitemShut
  {NoStop}%
\bibitem [{\citenamefont {Šupić}\ \emph {et~al.}(2023)\citenamefont
  {Šupić}, \citenamefont {Bowles}, \citenamefont {Renou}, \citenamefont
  {Acín},\ and\ \citenamefont {Hoban}}]{supic2022}%
  \BibitemOpen
  \bibfield  {author} {\bibinfo {author} {\bibfnamefont {I.}~\bibnamefont
  {Šupić}}, \bibinfo {author} {\bibfnamefont {J.}~\bibnamefont {Bowles}},
  \bibinfo {author} {\bibfnamefont {M.-O.}\ \bibnamefont {Renou}}, \bibinfo
  {author} {\bibfnamefont {A.}~\bibnamefont {Acín}},\ and\ \bibinfo {author}
  {\bibfnamefont {M.~J.}\ \bibnamefont {Hoban}},\ }\href
  {https://doi.org/10.1038/s41567-023-01945-4} {\bibfield  {journal} {\bibinfo
  {journal} {Nat. Phys.}\ } (\bibinfo {year} {2023})}\BibitemShut {NoStop}%
\bibitem [{\citenamefont {Renou}\ \emph {et~al.}(2019)\citenamefont {Renou},
  \citenamefont {B\"aumer}, \citenamefont {Boreiri}, \citenamefont {Brunner},
  \citenamefont {Gisin},\ and\ \citenamefont {Beigi}}]{renou2019genuine}%
  \BibitemOpen
  \bibfield  {author} {\bibinfo {author} {\bibfnamefont {M.-O.}\ \bibnamefont
  {Renou}}, \bibinfo {author} {\bibfnamefont {E.}~\bibnamefont {B\"aumer}},
  \bibinfo {author} {\bibfnamefont {S.}~\bibnamefont {Boreiri}}, \bibinfo
  {author} {\bibfnamefont {N.}~\bibnamefont {Brunner}}, \bibinfo {author}
  {\bibfnamefont {N.}~\bibnamefont {Gisin}},\ and\ \bibinfo {author}
  {\bibfnamefont {S.}~\bibnamefont {Beigi}},\ }\href
  {https://doi.org/10.1103/PhysRevLett.123.140401} {\bibfield  {journal}
  {\bibinfo  {journal} {Phys. Rev. Lett.}\ }\textbf {\bibinfo {volume} {123}},\
  \bibinfo {pages} {140401} (\bibinfo {year} {2019})}\BibitemShut {NoStop}%
\bibitem [{\citenamefont {Kriv\'achy}\ \emph {et~al.}(2020)\citenamefont
  {Kriv\'achy}, \citenamefont {Cai}, \citenamefont {Cavalcanti}, \citenamefont
  {Tavakoli}, \citenamefont {Gisin},\ and\ \citenamefont
  {Brunner}}]{krivachy2019neural}%
  \BibitemOpen
  \bibfield  {author} {\bibinfo {author} {\bibfnamefont {T.}~\bibnamefont
  {Kriv\'achy}}, \bibinfo {author} {\bibfnamefont {Y.}~\bibnamefont {Cai}},
  \bibinfo {author} {\bibfnamefont {D.}~\bibnamefont {Cavalcanti}}, \bibinfo
  {author} {\bibfnamefont {A.}~\bibnamefont {Tavakoli}}, \bibinfo {author}
  {\bibfnamefont {N.}~\bibnamefont {Gisin}},\ and\ \bibinfo {author}
  {\bibfnamefont {N.}~\bibnamefont {Brunner}},\ }\href
  {https://doi.org/10.1038/s41534-020-00305-x} {\bibfield  {journal} {\bibinfo
  {journal} {npj Quantum Inf.}\ }\textbf {\bibinfo {volume} {6}},\ \bibinfo
  {pages} {70} (\bibinfo {year} {2020})}\BibitemShut {NoStop}%
\bibitem [{\citenamefont {Gisin}\ \emph {et~al.}(2017)\citenamefont {Gisin},
  \citenamefont {Mei}, \citenamefont {Tavakoli}, \citenamefont {Renou},\ and\
  \citenamefont {Brunner}}]{Gisin2017}%
  \BibitemOpen
  \bibfield  {author} {\bibinfo {author} {\bibfnamefont {N.}~\bibnamefont
  {Gisin}}, \bibinfo {author} {\bibfnamefont {Q.}~\bibnamefont {Mei}}, \bibinfo
  {author} {\bibfnamefont {A.}~\bibnamefont {Tavakoli}}, \bibinfo {author}
  {\bibfnamefont {M.-O.}\ \bibnamefont {Renou}},\ and\ \bibinfo {author}
  {\bibfnamefont {N.}~\bibnamefont {Brunner}},\ }\href
  {https://doi.org/10.1103/PhysRevA.96.020304} {\bibfield  {journal} {\bibinfo
  {journal} {Phys. Rev. A}\ }\textbf {\bibinfo {volume} {96}},\ \bibinfo
  {pages} {020304(R)} (\bibinfo {year} {2017})}\BibitemShut {NoStop}%
\bibitem [{\citenamefont {Branciard}\ \emph {et~al.}(2012)\citenamefont
  {Branciard}, \citenamefont {Rosset}, \citenamefont {Gisin},\ and\
  \citenamefont {Pironio}}]{Branciard2012}%
  \BibitemOpen
  \bibfield  {author} {\bibinfo {author} {\bibfnamefont {C.}~\bibnamefont
  {Branciard}}, \bibinfo {author} {\bibfnamefont {D.}~\bibnamefont {Rosset}},
  \bibinfo {author} {\bibfnamefont {N.}~\bibnamefont {Gisin}},\ and\ \bibinfo
  {author} {\bibfnamefont {S.}~\bibnamefont {Pironio}},\ }\href
  {https://doi.org/10.1103/PhysRevA.85.032119} {\bibfield  {journal} {\bibinfo
  {journal} {Phys. Rev. A}\ }\textbf {\bibinfo {volume} {85}},\ \bibinfo
  {pages} {032119} (\bibinfo {year} {2012})}\BibitemShut {NoStop}%
\bibitem [{\citenamefont {Fritz}(2012)}]{Fritz2012}%
  \BibitemOpen
  \bibfield  {author} {\bibinfo {author} {\bibfnamefont {T.}~\bibnamefont
  {Fritz}},\ }\href {https://doi.org/10.1088/1367-2630/14/10/103001} {\bibfield
   {journal} {\bibinfo  {journal} {New J. Phys.}\ }\textbf {\bibinfo {volume}
  {14}},\ \bibinfo {pages} {103001} (\bibinfo {year} {2012})}\BibitemShut
  {NoStop}%
\bibitem [{\citenamefont {Pozas-Kerstjens}\ \emph {et~al.}(2022)\citenamefont
  {Pozas-Kerstjens}, \citenamefont {Gisin},\ and\ \citenamefont
  {Tavakoli}}]{pozas2021fullnn}%
  \BibitemOpen
  \bibfield  {author} {\bibinfo {author} {\bibfnamefont {A.}~\bibnamefont
  {Pozas-Kerstjens}}, \bibinfo {author} {\bibfnamefont {N.}~\bibnamefont
  {Gisin}},\ and\ \bibinfo {author} {\bibfnamefont {A.}~\bibnamefont
  {Tavakoli}},\ }\href {https://doi.org/10.1103/PhysRevLett.128.010403}
  {\bibfield  {journal} {\bibinfo  {journal} {Phys. Rev. Lett.}\ }\textbf
  {\bibinfo {volume} {128}},\ \bibinfo {pages} {010403} (\bibinfo {year}
  {2022})}\BibitemShut {NoStop}%
\bibitem [{\citenamefont {Šupić}\ \emph {et~al.}(2022)\citenamefont
  {Šupić}, \citenamefont {Bancal}, \citenamefont {Cai},\ and\ \citenamefont
  {Brunner}}]{supicGenuine}%
  \BibitemOpen
  \bibfield  {author} {\bibinfo {author} {\bibfnamefont {I.}~\bibnamefont
  {Šupić}}, \bibinfo {author} {\bibfnamefont {J.-D.}\ \bibnamefont {Bancal}},
  \bibinfo {author} {\bibfnamefont {Y.}~\bibnamefont {Cai}},\ and\ \bibinfo
  {author} {\bibfnamefont {N.}~\bibnamefont {Brunner}},\ }\href
  {https://doi.org/10.1103/PhysRevA.105.022206} {\bibfield  {journal} {\bibinfo
   {journal} {Phys. Rev. A}\ }\textbf {\bibinfo {volume} {105}},\ \bibinfo
  {pages} {022206} (\bibinfo {year} {2022})}\BibitemShut {NoStop}%
\bibitem [{\citenamefont {Renou}\ and\ \citenamefont
  {Beigi}(2022{\natexlab{a}})}]{Renou2020PRA}%
  \BibitemOpen
  \bibfield  {author} {\bibinfo {author} {\bibfnamefont {M.-O.}\ \bibnamefont
  {Renou}}\ and\ \bibinfo {author} {\bibfnamefont {S.}~\bibnamefont {Beigi}},\
  }\href {https://doi.org/10.1103/PhysRevA.105.022408} {\bibfield  {journal}
  {\bibinfo  {journal} {Phys. Rev. A}\ }\textbf {\bibinfo {volume} {105}},\
  \bibinfo {pages} {022408} (\bibinfo {year} {2022}{\natexlab{a}})}\BibitemShut
  {NoStop}%
\bibitem [{\citenamefont {Renou}\ and\ \citenamefont
  {Beigi}(2022{\natexlab{b}})}]{Renou2020PRL}%
  \BibitemOpen
  \bibfield  {author} {\bibinfo {author} {\bibfnamefont {M.-O.}\ \bibnamefont
  {Renou}}\ and\ \bibinfo {author} {\bibfnamefont {S.}~\bibnamefont {Beigi}},\
  }\href {https://doi.org/10.1103/PhysRevLett.128.060401} {\bibfield  {journal}
  {\bibinfo  {journal} {Phys. Rev. Lett.}\ }\textbf {\bibinfo {volume} {128}},\
  \bibinfo {pages} {060401} (\bibinfo {year} {2022}{\natexlab{b}})}\BibitemShut
  {NoStop}%
\bibitem [{\citenamefont {Abiuso}\ \emph {et~al.}(2022)\citenamefont {Abiuso},
  \citenamefont {Kriv\'achy}, \citenamefont {Boghiu}, \citenamefont {Renou},
  \citenamefont {Pozas-Kerstjens},\ and\ \citenamefont
  {Ac\'{\i}n}}]{abiuso2021quantum}%
  \BibitemOpen
  \bibfield  {author} {\bibinfo {author} {\bibfnamefont {P.}~\bibnamefont
  {Abiuso}}, \bibinfo {author} {\bibfnamefont {T.}~\bibnamefont {Kriv\'achy}},
  \bibinfo {author} {\bibfnamefont {E.-C.}\ \bibnamefont {Boghiu}}, \bibinfo
  {author} {\bibfnamefont {M.-O.}\ \bibnamefont {Renou}}, \bibinfo {author}
  {\bibfnamefont {A.}~\bibnamefont {Pozas-Kerstjens}},\ and\ \bibinfo {author}
  {\bibfnamefont {A.}~\bibnamefont {Ac\'{\i}n}},\ }\href
  {https://doi.org/10.1103/PhysRevResearch.4.L012041} {\bibfield  {journal}
  {\bibinfo  {journal} {Phys. Rev. Research}\ }\textbf {\bibinfo {volume}
  {4}},\ \bibinfo {pages} {L012041} (\bibinfo {year} {2022})}\BibitemShut
  {NoStop}%
\bibitem [{\citenamefont {Wolfe}\ \emph {et~al.}(2019)\citenamefont {Wolfe},
  \citenamefont {Spekkens},\ and\ \citenamefont {Fritz}}]{wolfe2019inflation}%
  \BibitemOpen
  \bibfield  {author} {\bibinfo {author} {\bibfnamefont {E.}~\bibnamefont
  {Wolfe}}, \bibinfo {author} {\bibfnamefont {R.~W.}\ \bibnamefont
  {Spekkens}},\ and\ \bibinfo {author} {\bibfnamefont {T.}~\bibnamefont
  {Fritz}},\ }\href {https://doi.org/10.1515/jci-2017-0020} {\bibfield
  {journal} {\bibinfo  {journal} {J. Causal Inference}\ }\textbf {\bibinfo
  {volume} {7}},\ \bibinfo {pages} {20170020} (\bibinfo {year}
  {2019})}\BibitemShut {NoStop}%
\bibitem [{\citenamefont {Chaves}\ \emph {et~al.}(2015)\citenamefont {Chaves},
  \citenamefont {Majenz},\ and\ \citenamefont {Gross}}]{Chaves2015}%
  \BibitemOpen
  \bibfield  {author} {\bibinfo {author} {\bibfnamefont {R.}~\bibnamefont
  {Chaves}}, \bibinfo {author} {\bibfnamefont {C.}~\bibnamefont {Majenz}},\
  and\ \bibinfo {author} {\bibfnamefont {D.}~\bibnamefont {Gross}},\ }\href
  {https://doi.org/10.1038/ncomms6766} {\bibfield  {journal} {\bibinfo
  {journal} {Nat. Commun.}\ }\textbf {\bibinfo {volume} {6}},\ \bibinfo {pages}
  {5766} (\bibinfo {year} {2015})}\BibitemShut {NoStop}%
\bibitem [{\citenamefont {Weilenmann}\ and\ \citenamefont
  {Colbeck}(2017)}]{Weilenmann2017}%
  \BibitemOpen
  \bibfield  {author} {\bibinfo {author} {\bibfnamefont {M.}~\bibnamefont
  {Weilenmann}}\ and\ \bibinfo {author} {\bibfnamefont {R.}~\bibnamefont
  {Colbeck}},\ }\href {https://doi.org/10.1098/rspa.2017.0483} {\bibfield
  {journal} {\bibinfo  {journal} {Proc. R. Soc. A}\ }\textbf {\bibinfo {volume}
  {473}},\ \bibinfo {pages} {20170483} (\bibinfo {year} {2017})}\BibitemShut
  {NoStop}%
\bibitem [{\citenamefont {Pozas-Kerstjens}\ \emph {et~al.}(2019)\citenamefont
  {Pozas-Kerstjens}, \citenamefont {Rabelo}, \citenamefont {Rudnicki},
  \citenamefont {Chaves}, \citenamefont {Cavalcanti}, \citenamefont
  {Navascu\'es},\ and\ \citenamefont {Ac\'{\i}n}}]{Pozas2019}%
  \BibitemOpen
  \bibfield  {author} {\bibinfo {author} {\bibfnamefont {A.}~\bibnamefont
  {Pozas-Kerstjens}}, \bibinfo {author} {\bibfnamefont {R.}~\bibnamefont
  {Rabelo}}, \bibinfo {author} {\bibfnamefont {{\L}.}~\bibnamefont {Rudnicki}},
  \bibinfo {author} {\bibfnamefont {R.}~\bibnamefont {Chaves}}, \bibinfo
  {author} {\bibfnamefont {D.}~\bibnamefont {Cavalcanti}}, \bibinfo {author}
  {\bibfnamefont {M.}~\bibnamefont {Navascu\'es}},\ and\ \bibinfo {author}
  {\bibfnamefont {A.}~\bibnamefont {Ac\'{\i}n}},\ }\href
  {https://doi.org/10.1103/PhysRevLett.123.140503} {\bibfield  {journal}
  {\bibinfo  {journal} {Phys. Rev. Lett.}\ }\textbf {\bibinfo {volume} {123}},\
  \bibinfo {pages} {140503} (\bibinfo {year} {2019})}\BibitemShut {NoStop}%
\bibitem [{\citenamefont {Fraser}\ and\ \citenamefont
  {Wolfe}(2018)}]{Fraser2018}%
  \BibitemOpen
  \bibfield  {author} {\bibinfo {author} {\bibfnamefont {T.~C.}\ \bibnamefont
  {Fraser}}\ and\ \bibinfo {author} {\bibfnamefont {E.}~\bibnamefont {Wolfe}},\
  }\href {https://doi.org/10.1103/PhysRevA.98.022113} {\bibfield  {journal}
  {\bibinfo  {journal} {Phys. Rev. A}\ }\textbf {\bibinfo {volume} {98}},\
  \bibinfo {pages} {022113} (\bibinfo {year} {2018})}\BibitemShut {NoStop}%
\bibitem [{\citenamefont {Navascu\'{e}s}\ and\ \citenamefont
  {Wolfe}(2020)}]{navascues2017convergence}%
  \BibitemOpen
  \bibfield  {author} {\bibinfo {author} {\bibfnamefont {M.}~\bibnamefont
  {Navascu\'{e}s}}\ and\ \bibinfo {author} {\bibfnamefont {E.}~\bibnamefont
  {Wolfe}},\ }\href {https://doi.org/10.1515/jci-2018-0008} {\bibfield
  {journal} {\bibinfo  {journal} {J. Causal Inference}\ }\textbf {\bibinfo
  {volume} {8}},\ \bibinfo {pages} {70 } (\bibinfo {year} {2020})}\BibitemShut
  {NoStop}%
\bibitem [{\citenamefont {Pozas-Kerstjens}(2019)}]{alexThesis}%
  \BibitemOpen
  \bibfield  {author} {\bibinfo {author} {\bibfnamefont {A.}~\bibnamefont
  {Pozas-Kerstjens}},\ }\emph {\bibinfo {title} {Quantum information outside
  quantum information}},\ \href {http://hdl.handle.net/10803/667696} {Ph.D.
  thesis} (\bibinfo {year} {2019})\BibitemShut {NoStop}%
\bibitem [{\citenamefont {Boreiri}\ \emph {et~al.}(2022)\citenamefont
  {Boreiri}, \citenamefont {Girardin}, \citenamefont {Ulu}, \citenamefont
  {Lypka-Bartosik}, \citenamefont {Brunner},\ and\ \citenamefont
  {Sekatski}}]{Boreiri2022}%
  \BibitemOpen
  \bibfield  {author} {\bibinfo {author} {\bibfnamefont {S.}~\bibnamefont
  {Boreiri}}, \bibinfo {author} {\bibfnamefont {A.}~\bibnamefont {Girardin}},
  \bibinfo {author} {\bibfnamefont {B.}~\bibnamefont {Ulu}}, \bibinfo {author}
  {\bibfnamefont {P.}~\bibnamefont {Lypka-Bartosik}}, \bibinfo {author}
  {\bibfnamefont {N.}~\bibnamefont {Brunner}},\ and\ \bibinfo {author}
  {\bibfnamefont {P.}~\bibnamefont {Sekatski}},\ }\href
  {https://doi.org/10.48550/ARXIV.2207.08532} {\bibinfo {title} {Towards a
  minimal example of quantum nonlocality without inputs}} (\bibinfo {year}
  {2022}),\ \Eprint {https://arxiv.org/abs/2207.08532} {arXiv:2207.08532}
  \BibitemShut {NoStop}%
\bibitem [{\citenamefont {G\"artner}\ and\ \citenamefont
  {Matou\v{s}ek}(2007)}]{FarkasLemma}%
  \BibitemOpen
  \bibfield  {author} {\bibinfo {author} {\bibfnamefont {B.}~\bibnamefont
  {G\"artner}}\ and\ \bibinfo {author} {\bibfnamefont {J.}~\bibnamefont
  {Matou\v{s}ek}},\ }\href {https://doi.org/10.1007/978-3-540-30717-4} {\emph
  {\bibinfo {title} {Understanding and using linear programming}}}\ (\bibinfo
  {publisher} {Springer},\ \bibinfo {year} {2007})\BibitemShut {NoStop}%
\bibitem [{\citenamefont {Baccari}\ \emph {et~al.}(2017)\citenamefont
  {Baccari}, \citenamefont {Cavalcanti}, \citenamefont {Wittek},\ and\
  \citenamefont {Ac\'{\i}n}}]{baccari2017detection}%
  \BibitemOpen
  \bibfield  {author} {\bibinfo {author} {\bibfnamefont {F.}~\bibnamefont
  {Baccari}}, \bibinfo {author} {\bibfnamefont {D.}~\bibnamefont {Cavalcanti}},
  \bibinfo {author} {\bibfnamefont {P.}~\bibnamefont {Wittek}},\ and\ \bibinfo
  {author} {\bibfnamefont {A.}~\bibnamefont {Ac\'{\i}n}},\ }\href
  {https://doi.org/10.1103/PhysRevX.7.021042} {\bibfield  {journal} {\bibinfo
  {journal} {Phys. Rev. X}\ }\textbf {\bibinfo {volume} {7}},\ \bibinfo {pages}
  {021042} (\bibinfo {year} {2017})}\BibitemShut {NoStop}%
\bibitem [{\citenamefont {Pozas-Kerstjens}(2022)}]{compAppendix}%
  \BibitemOpen
  \bibfield  {author} {\bibinfo {author} {\bibfnamefont {A.}~\bibnamefont
  {Pozas-Kerstjens}},\ }\href {https://doi.org/10.5281/zenodo.6406666}
  {\bibfield  {journal} {\bibinfo  {journal} {Zenodo}\ }\textbf {\bibinfo
  {volume} {6406666}} (\bibinfo {year} {2022})}\BibitemShut {NoStop}%
\bibitem [{\citenamefont {Bancal}\ and\ \citenamefont
  {Gisin}(2021)}]{Bancal2021}%
  \BibitemOpen
  \bibfield  {author} {\bibinfo {author} {\bibfnamefont {J.-D.}\ \bibnamefont
  {Bancal}}\ and\ \bibinfo {author} {\bibfnamefont {N.}~\bibnamefont {Gisin}},\
  }\href {https://doi.org/10.1103/PhysRevA.104.052212} {\bibfield  {journal}
  {\bibinfo  {journal} {Phys. Rev. A}\ }\textbf {\bibinfo {volume} {104}},\
  \bibinfo {pages} {052212} (\bibinfo {year} {2021})}\BibitemShut {NoStop}%
\bibitem [{\citenamefont {Carter}\ and\ \citenamefont
  {Nielsen}(2017)}]{carter2017AIA}%
  \BibitemOpen
  \bibfield  {author} {\bibinfo {author} {\bibfnamefont {S.}~\bibnamefont
  {Carter}}\ and\ \bibinfo {author} {\bibfnamefont {M.}~\bibnamefont
  {Nielsen}},\ }\href {https://doi.org/10.23915/distill.00009} {\bibfield
  {journal} {\bibinfo  {journal} {Distill}\ }\textbf {\bibinfo {volume} {9}}
  (\bibinfo {year} {2017})}\BibitemShut {NoStop}%
\bibitem [{\citenamefont {Venderley}\ \emph {et~al.}(2018)\citenamefont
  {Venderley}, \citenamefont {Khemani},\ and\ \citenamefont
  {Kim}}]{venderley2018}%
  \BibitemOpen
  \bibfield  {author} {\bibinfo {author} {\bibfnamefont {J.}~\bibnamefont
  {Venderley}}, \bibinfo {author} {\bibfnamefont {V.}~\bibnamefont {Khemani}},\
  and\ \bibinfo {author} {\bibfnamefont {E.-A.}\ \bibnamefont {Kim}},\ }\href
  {https://doi.org/10.1103/PhysRevLett.120.257204} {\bibfield  {journal}
  {\bibinfo  {journal} {Phys. Rev. Lett.}\ }\textbf {\bibinfo {volume} {120}},\
  \bibinfo {pages} {257204} (\bibinfo {year} {2018})}\BibitemShut {NoStop}%
\bibitem [{\citenamefont {Kottmann}\ \emph {et~al.}(2020)\citenamefont
  {Kottmann}, \citenamefont {Huembeli}, \citenamefont {Lewenstein},\ and\
  \citenamefont {Ac\'{\i}n}}]{kottmann2020}%
  \BibitemOpen
  \bibfield  {author} {\bibinfo {author} {\bibfnamefont {K.}~\bibnamefont
  {Kottmann}}, \bibinfo {author} {\bibfnamefont {P.}~\bibnamefont {Huembeli}},
  \bibinfo {author} {\bibfnamefont {M.}~\bibnamefont {Lewenstein}},\ and\
  \bibinfo {author} {\bibfnamefont {A.}~\bibnamefont {Ac\'{\i}n}},\ }\href
  {https://doi.org/10.1103/PhysRevLett.125.170603} {\bibfield  {journal}
  {\bibinfo  {journal} {Phys. Rev. Lett.}\ }\textbf {\bibinfo {volume} {125}},\
  \bibinfo {pages} {170603} (\bibinfo {year} {2020})}\BibitemShut {NoStop}%
\end{thebibliography}%

\onecolumngrid
\appendix
\setcounter{figure}{0}
\renewcommand{\thefigure}{S\arabic{figure}}

\section{Conditions on $q$ and the feasible polytope}
\label{app:polytope}
As an introduction, let us begin by analyzing which are the necessary conditions for Eq.~\eqref{eq:q} in the main text to describe a valid probability distribution.
A general four-partite, binary-outcome distribution with outcomes taking values $\pm1$ reads
\begin{align}
  q(i,j,k,t) = \frac{1}{2^4}&\left[1 + i E_A + j E_B + k E_C+ t F + ij E_{AB} + jk E_{BC} + ki E_{AC} + t (i F_{A} + j F_{B} + k F_{C}) \right. \notag\\
  & \left. + ijk E_{ABC} + t (ij F_{AB} + jk F_{BC} + ki F_{AC}) + ijkt F_{ABC}\right],
\end{align}
where the parameters $\{E_A,E_B,E_C,E_{AB},E_{BC},E_{AC},E_{ABC},F,F_A,F_B,F_C,F_{AB},F_{BC},F_{AC},F_{ABC}\}\,{\in}\,[-1,1]$ characterize all possible distributions.
Next, some of these parameters can be fixed from identifications with $P_u(a,b,c)$.
The fact that the hidden variables of the coarse-grained distribution are uniformly random translates into $q(t{=}-1)\,{=}\,q(t{=}1)$, which fixes $F\,{=}\,0$.
In a similar manner, the constraints on single-party values \cite[Eq. (8)]{renou2019genuine} imply $E_A\,{=}\,E_B\,{=}\,E_C\,{=}\,0$ and $F_A\,{=}\,F_B\,{=}\,F_C\,{=}\,u^2-v^2$ (where, recall, $v\,{=}\,\sqrt{1-u^2}$), and the constraints on three-party values \cite[Eq. (7)]{renou2019genuine} imply $E_{AB}\,{=}\,E_{BC}\,{=}\,E_{AC}\,{=}\,(u^2-v^2)^2$ and $E_{ABC}\,{=}\,8u^3v^3$.
We are thus left with a general parametrization of $q(i,j,k,t)$ that depends only on the four parameters $F_{AB}$, $F_{BC}$, $F_{AC}$ and $F_{ABC}$ and reads
\begin{align}
  q_u(i,j,k,t) = \frac{1}{2^4}&\left[1 + (u^2-v^2)^2(ij+jk+ki) + (u^2-v^2)(i+j+k)t + 8u^3v^3ijk\right. \notag\\
	& \left.\,+ \,t(ij F_{AB} + jk F_{BC} + ki F_{AC}) + ijkt F_{ABC}\right].
\end{align}

\begin{figure}
	\includegraphics[width=0.27\textwidth]{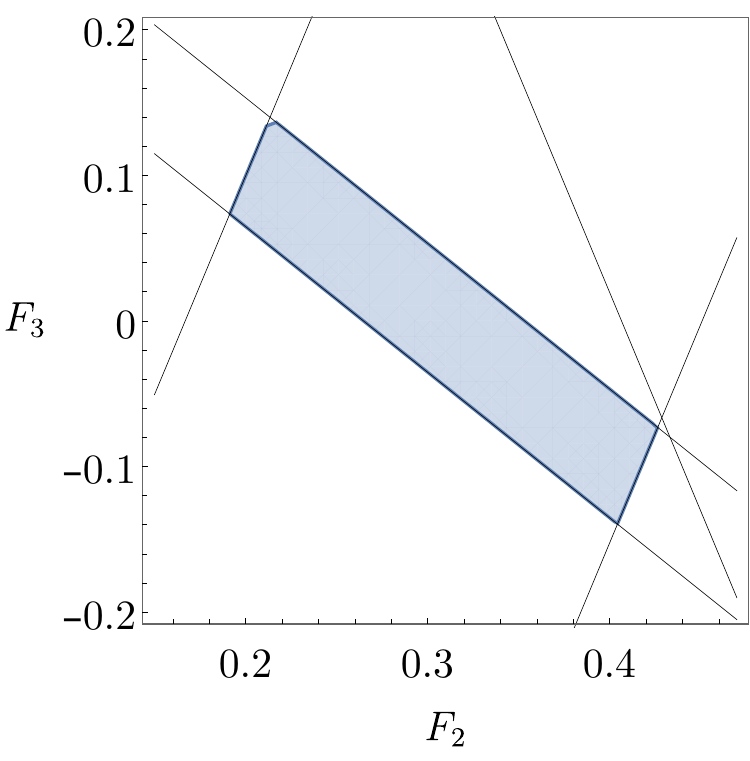}
	\caption{Polytope of values that produce well-defined probability distributions $q_u(i,j,k,t)$, defined by Eqs.~\eqref{eq:polytope}, for $u\,{=}\,0.8090$ and projected onto $(F_{AB}\,{=}\,F_{BC}\,{=}\,F_{AC}\,{\equiv}\, F_2, F_{ABC}\,{\equiv}\, F_3)$.
	For increasing $u$, the line not defining the quadrilateral moves downwards, intersecting with it and eventually producing an empty region for $u\,{=}\,\sqrt{\frac{-3 + (9 + 6 \sqrt{3})^{2/3}}{2 (9 + 6 \sqrt{3})^{1/3}}}\,{\approx}\, 0.8860$.
	For decreasing $u$, the area of the quadrilateral reduces, eventually producing a point at $u\,{=}\,\frac{1}{\sqrt{2}}$ and an empty region for lower values.}
	\label{fig:polytope}
\end{figure}

In order for $q(i,j,k,t)$ to be well defined, all probabilities must be positive.
These conditions give rise to a polytope in the space $(F_{AB},F_{BC},F_{AC},F_{ABC})$, determined by the following sixteen linear inequalities:
\begin{subequations}
	\begin{align}
		1 - F_{AB} - F_{AC} - F_{BC} + F_{ABC} + 6 u^2 (u^2 - v^2) - 8 u^3 v^3 &\geq 0, \\
		1 + F_{AB} + F_{AC} + F_{BC} - F_{ABC} - 6 v^2 (u^2 - v^2) - 8 u^3 v^3 &\geq 0, \label{eq:polytope:v1} \\
		1 - F_{AB} + F_{AC} + F_{BC} - F_{ABC} + 2 v^2 (u^2 - v^2) + 8 u^3 v^3 &\geq 0, \\
		1 + F_{AB} - F_{AC} - F_{BC} + F_{ABC} - 2 u^2 (u^2 - v^2) + 8 u^3 v^3 &\geq 0, \\
		1 + F_{AB} - F_{AC} + F_{BC} - F_{ABC} + 2 v^2 (u^2 - v^2) + 8 u^3 v^3 &\geq 0, \\
		1 - F_{AB} + F_{AC} - F_{BC} + F_{ABC} - 2 u^2 (u^2 - v^2) + 8 u^3 v^3 &\geq 0, \\
		1 + F_{AB} + F_{AC} - F_{BC} + F_{ABC} - 2 u^2 (u^2 - v^2) - 8 u^3 v^3 &\geq 0, \\
		1 - F_{AB} - F_{AC} + F_{BC} - F_{ABC} + 2 v^2 (u^2 - v^2) - 8 u^3 v^3 &\geq 0, \label{eq:polytope:v2} \\
		1 + F_{AB} + F_{AC} - F_{BC} - F_{ABC} + 2 v^2 (u^2 - v^2) + 8 u^3 v^3 &\geq 0, \\
		1 - F_{AB} - F_{AC} + F_{BC} + F_{ABC} - 2 u^2 (u^2 - v^2) + 8 u^3 v^3 &\geq 0, \\
		1 + F_{AB} - F_{AC} + F_{BC} + F_{ABC} - 2 u^2 (u^2 - v^2) - 8 u^3 v^3 &\geq 0, \\
		1 - F_{AB} + F_{AC} - F_{BC} - F_{ABC} + 2 v^2 (u^2 - v^2) - 8 u^3 v^3 &\geq 0, \label{eq:polytope:v3} \\
		1 - F_{AB} + F_{AC} + F_{BC} + F_{ABC} - 2 u^2 (u^2 - v^2) - 8 u^3 v^3 &\geq 0, \\
		1 + F_{AB} - F_{AC} - F_{BC} - F_{ABC} + 2 v^2 (u^2 - v^2) - 8 u^3 v^3 &\geq 0, \label{eq:polytope:v4} \\
		1 - F_{AB} - F_{AC} - F_{BC} - F_{ABC} - 6 v^2 (u^2 - v^2) + 8 u^3 v^3 &\geq 0, \\
		1 + F_{AB} + F_{AC} + F_{BC} + F_{ABC} + 6 u^2 (u^2 - v^2) + 8 u^3 v^3 &\geq 0.
	\end{align}
	\label{eq:polytope}%
\end{subequations}
An illustration of the projection of the polytope for $u\,{=}\,0.8090$ onto the subspace satisfying $F_{AB}\,{=}\,F_{BC}\,{=}\,F_{AC}$ is depicted in Fig.~\ref{fig:polytope}.
A rather lengthy but straightforward calculation shows that the polytope is empty for $u\,{>}\, \sqrt{\frac{-3 + (9 + 6 \sqrt{3})^{2/3}}{2 (9 + 6 \sqrt{3})^{1/3}}}\,{\equiv}\, u_0$.
This recovers the proof developed in Ref.~\cite{renou2019genuine}.
Moreover, similar calculations allow to show that the polytope becomes a single point at both $u=u_0$ and $u\,{=}\,1/\sqrt{2}$.
For the former, an explicit construction is derived in \cite[Appendix D]{renou2019genuine}.
For the latter, it is easy to see that $P_u(a,b,c)$ admits a model of the form of Eq.~\eqref{eq:trianglelocal} in the main text.

\section{Equivalence between triangle-locality of $P_u(a,b,c)$ and existence of a suitable $q_u(i,j,k,t)$}
\label{app:equiv}
Here, we prove that $P_u(a,b,c)$ is triangle local if and only if there exists at least one distribution $q_u(i,j,k,t)$ satisfying the positivity conditions given in Section~\ref{app:polytope} and such that both $q_u^+(i,j,k)$ and $q_u^-(i,j,k)$ are triangle local.

\subsection{$P_u(a,b,c)$ is triangle local $\Rightarrow \exists\, q_u(i,j,k,t)$ compatible such that $q_u^+(i,j,k)$ and $q_u^-(i,j,k)$ are triangle local}
Let us begin assuming that $P_u(a,b,c)$ is triangle local and obtained with a model (or strategy) called $\mathcal{S}$.
This model determines a coarse-grained model, $\mathcal{S}_\mathrm{TC}$, when grouping the outputs $\bar{1}_+,\bar{1}_-$ into a single output, $\bar{1}$, that reproduces the coarse-grained distribution $P_{\mathrm{TC}}(a,b,c)$.

References~\cite{renou2019genuine,Renou2020PRA,Renou2020PRL} proved that there is essentially a unique way to obtain $P_{\mathrm{TC}}(a,b,c)$ with a triangle-local model, called the Token-Counting strategy.
In this strategy for $P_{\mathrm{TC}}(a,b,c)$, each source distributes a uniformly binary random hidden variable (a token) and each party produces her output $\{\bar{0},\bar{1},\bar{2}\}$ by counting the total number of tokens received.

This implies that in $\mathcal{S}$, the sources $\alpha$, $\beta$, and $\gamma$, can be taken to each distribute a uniformly random bit (call them $T_\alpha(\alpha)$, $T_\beta(\beta)$ and $T_\gamma(\gamma)$) that ultimately determine whether a given party outputs $\bar{0}$, $\bar{2}$, or something else, and additional information that specify, in the case of outputting something else, whether that is $\bar{1}_+$ or $\bar{1}_-$.
For instance, to produce her output, Alice first looks at the values of $T_\beta(\beta)$ and $T_\gamma(\gamma)$.
If these variables indicate that she receives $0$ or $2$ tokens, she outputs $\bar{0}$ or $\bar{2}$ accordingly.
Otherwise, she selects $\bar{1}_+$ or $\bar{1}_-$ depending on whether the token was received from the source $\beta$ or from the source $\gamma$, as well as the additional information sent by those two sources.

Now, consider the case in which the outputs of all parties are in $\{\bar{1}_+,\bar{1}_-\}$, which happens with probability $1/4$.
This case can be obtained in two ways: either Alice received her token from $\gamma$, Bob received his token from $\alpha$, and Charlie received his token from $\beta$ (we label this situation with $t\,{=}\,+1$), or Alice received her token from $\beta$, Bob received his token from $\gamma$, and Charlie received his token from $\alpha$ (which we label with $t\,{=}\,-1$).
One can thus introduce $q_u(i,j,k,t)$ to be the joint probability distribution of Alice, Bob and Charlie outputting $\bar{1}_i,\bar{1}_j,\bar{1}_k$ and the token taking the value $t$.

This distribution, because it is generated from the strategy $\mathcal{S}$, satisfies the constraints described in Appendix~\ref{app:polytope}.
Clearly, $q_u(t{=}{-}1)\,{=}\,q_u(t{=}{+}1)$ because in $\mathcal{S}_{TC}$ the tokens are uniformly distributed.
Moreover, the three-party marginal equation \cite[Eq. (7)]{renou2019genuine} is clearly satisfied.
Also, the single-party marginal equations \cite[Eq. (8)]{renou2019genuine} (e.g. $q_u(i,t{=}{+}1)\,{=}\,4P_u(\bar{1}_i,\bar{2},\bar{0})$) can be proven by paralleling Refs.~\cite{renou2019genuine,Renou2020PRA,Renou2020PRL}, exploiting the property that Alice's output is independent of the information distributed by the source $\alpha$, and similarly for the other parties.
Hence, $q_u(i,j,k,t)$ is compatible with the conditions of Appendix~\ref{app:polytope}.

Finally, both $q_u^+(i,j,k)$ and $q_u^-(i,j,k)$ admit triangle-local models.
Indeed, $q_u^+(i,j,k)$ can be obtained with the strategy $\mathcal{S}_+$ in which Alice, Bob and Charlie produce their outputs by always assuming that $t\,{=}\,+1$ and using the remaining information sent by the sources in order to produce either $\bar{1}_+$ or $\bar{1}_-$.
$q_u^-(i,j,k)$ can be obtained similarly, considering $t\,{=}\,-1$ instead.

\subsection{$P_u(a,b,c)$ is triangle local $\Leftarrow \exists\, q_u(i,j,k,t)$ compatible such that $q_u^+(i,j,k), q_u^-(i,j,k)$ are triangle local}
Assume now that there exists one distribution $q_u(i,j,k,t)$ (i.e., one particular value of $F_{AB}$, $F_{BC}$, $F_{AC}$ and $F_{ABC}$) compatible with the constraints given in Appendix~\ref{app:polytope} such that both $q_u^+(i,j,k)$ and $q_u^-(i,j,k)$ are triangle local.
This, is, that there exist variables $\alpha_+$, $\beta_+$ and $\gamma_+$ that produce $q_u^+(i,j,k)$ via Eq.~\eqref{eq:trianglelocal} in the main text (let us call this strategy $\mathcal{S}_+$), and variables $\alpha_-$, $\beta_-$, $\gamma_-$ that produce $q_u^-(i,j,k)$ (we denote this strategy as $\mathcal{S}_-$).
From these, we will find a triangle-local strategy $\mathcal{S}$ for $P_u(a,b,c)$.

The strategy is constructed in the following way.
First, take the source $\alpha$ to distribute $(\alpha_+,\alpha_-,T_\alpha)$, this is, the information of both previous models ($\alpha_+$ and $\alpha_-$), and an additional uniformly binary random variable that will denote whether a token is sent either to Bob or to Charlie.
The sources $\beta$ and $\gamma$ are constructed in a similar way.
Then, when Alice receives the information from the sources she is connected to, namely $(\beta_+,\beta_-,T_\beta)$ and $(\gamma_+,\gamma_-,T_\gamma)$, she:
\begin{enumerate}
\item Counts the total number of tokens $T_A=f(T_\beta,T_\gamma)\,{\in}\,\{\bar{0}, \bar{1}, \bar{2}\}$ she received,
\item Outputs $T_A$ if $T_A\,{\in}\,\{\bar{0},\bar{2}\}$,
\item Otherwise (i.e., when $T_A\,{=}\,\bar{1}$), she looks at where the token is coming from. If it comes from the source $\beta$, she uses the values of $\beta_+$ and $\gamma_+$ to produce her output according to the strategy $\mathcal{S}_+$. If, on the contrary, it comes from the source $\gamma$, she uses the values of $\beta_-$ and $\gamma_-$ to output according to $\mathcal{S}_-$.
\end{enumerate}
Bob and Charlie, on their ends, adopt analogous strategies.

The strategy outlined above, which we call $\mathcal{S}$, is indeed a triangle-local model for $P_u(a,b,c)$.
Consider first the probability to observe $a\,{=}\,\bar{1}_i$, $b\,{=}\,\bar{2}$, $c\,{=}\,\bar{0}$ in $\mathcal{S}$.
This is obtained when both $T_\alpha$ and $T_\gamma$ sends their tokens to Bob, $T_\beta$ sends its token to Alice, and Alice plays strategy $\mathcal{S}_+$ and obtains output $a\,{=}\,\bar{1}_i$.
Hence, it is obtained with probability $\frac{1}{8} q_u(i|t{=}{+}1)\,{=}\,\frac{1}{2} q(i,t{=}{+}1)\,{=}\,P_u(\bar{1}_i,\bar{2},\bar{0})$, where the last equality comes from the fact that $q(i,j,k,t)$ is compatible with the constraints given in Appendix~\ref{app:polytope}, and hence satisfies \cite[Eq. (8)]{renou2019genuine}.
Similarly, $\frac{1}{8} q_u(i|t{=}{-}1)\,{=}\,P_u(\bar{1}_i,\bar{0},\bar{2})$, $\frac{1}{8}q_u(j|t{=}{+}1)\,{=}\,P_u(\bar{0},\bar{1}_j,\bar{2})$, and so on.

Finally, consider the probability to observe $a\,{=}\,\bar{1}_i$, $b\,{=}\,\bar{1}_j$ and $c\,{=}\,\bar{1}_k$.
This is obtained when either $T_\alpha$ sends its token to Bob, $T_\beta$ sends its token to Charlie, $T_\gamma$ sends its token to Alice (which is labelled as $t\,{=}\,+1$) and all parties play strategy $\mathcal{S}_+$, or when $T_\alpha$ sends its token to Charlie, $T_\beta$ sends its token to Alice, $T_\gamma$ sends its tokens to Bob (which is labelled as $t\,{=}\,-1$) and all parties play strategy $\mathcal{S}_-$.
Hence, the outcome $a\,{=}\,\bar{1}_i$, $b\,{=}\,\bar{1}_j$, $c\,{=}\,\bar{1}_k$ is obtained with probability $\frac{1}{8}[q_u(i,j,k|t{=}{+}1)+q_u(i,j,k|t{=}{-}1)]\,{=}\,\frac{1}{2}[q_u(i,j,k,t{=}{+}1)+q_u(i,j,k,t{=}{-}1)]\,{=}\,\frac{1}{2}q_u(i,j,k)\,{=}\,P_u(\bar{1}_i,\bar{1}_j,\bar{1}_k)$, where the last equality comes from the fact that $q(i,j,k,t)$ is compatible with the constraints given in Appendix~\ref{app:polytope} hence satisfies \cite[Eq. (7)]{renou2019genuine}.
The remaining probabilities are zero in $P_u(a,b,c)$ and in the strategies $\mathcal{S}_+$ and $\mathcal{S}_-$, and thus the proof concludes.

\section{A triangle-local model for $q_u^+(i,j,k)$}
\label{app:qplusmodel}
Here we prove that, in a large range of $u$, there exists one choice of $F_{AB}$, $F_{BC}$, $F_{AC}$ and $F_{ABC}$ within the polytope of Eq.~\eqref{eq:polytope} for which $q^+_u(i,j,k)\,{\coloneqq}\, q_u(i,j,k|t{=}{+}1)$ admits a triangle-local model.
This choice is the intersection of Eqs.~\eqref{eq:polytope:v1}, \eqref{eq:polytope:v2}, \eqref{eq:polytope:v3} and \eqref{eq:polytope:v4}, which is described by $F_{AB}\,{=}\,F_{BC}\,{=}\,F_{AC}\,{=}\,2 v^2 (u^2 - v^2)$ and $F_{ABC}\,{=}\,1-8u^3v^3$, and gives rise to the distribution
\begin{subequations}
	\begin{align}
		q^{+}_u(+1,+1,-1)&=\frac{v^2}{2},\\
		q^{+}_u(-1,-1,-1)&=u^2-\frac{v^2}{2},
	\end{align}
	\label{eq:qplusmodel}%
\end{subequations}
plus permutations of parties, and all remaining probabilities being zero.
This distribution can be realized in a triangle-local model as follows: set the hidden variables $\alpha$, $\beta$ and $\gamma$ to be binary, with equal probability distributions where any one value is set to be $(1 + \sqrt{u^2-v^2})/2$.
If every party outputs $-1$ when the received hidden variables are equal and $+1$ otherwise, the distribution \eqref{eq:qplusmodel} is recovered.

The point $(F_{AB}, F_{BC}, F_{AC}, F_{ABC})$ described by the intersection of Eqs.~\eqref{eq:polytope:v1}, \eqref{eq:polytope:v2}, \eqref{eq:polytope:v3} and \eqref{eq:polytope:v4} is part of the polytope for all $1/\sqrt{2}\,{\leq}\, u \,{\leq}\, 1/2 \sqrt{1 + \sqrt{3^{4/3} + 3^{2/3}-5} + \sqrt{20/\sqrt{3^{4/3} + 3^{2/3} - 5} -10 - 3^{4/3} - 3^{2/3}}}\,{\approx}\,0.8457$, so at least for this whole range $q^+_u(i,j,k)$ cannot be used for proving the triangle nonlocality of $P_u(a,b,c)$.

\section{Inflation used and constraints imposed}
\label{app:inflation}
Inflation, introduced by Wolfe et al.~\cite{wolfe2019inflation}, is a powerful concept that enables the analysis of correlations in arbitrary causal structures, therefore including networks.
Its underlying mechanism is proving through contradiction: in the context of networks, if a distribution can be generated between some parties by using certain sources, then one can consider which kinds of distributions one would be able to generate if given access to multiple copies of such parties and sources.
The networks that are generated by arranging those copies are called inflations of the original network.

\begin{figure*}[h]
	\centering
	\includegraphics[width=0.3\columnwidth]{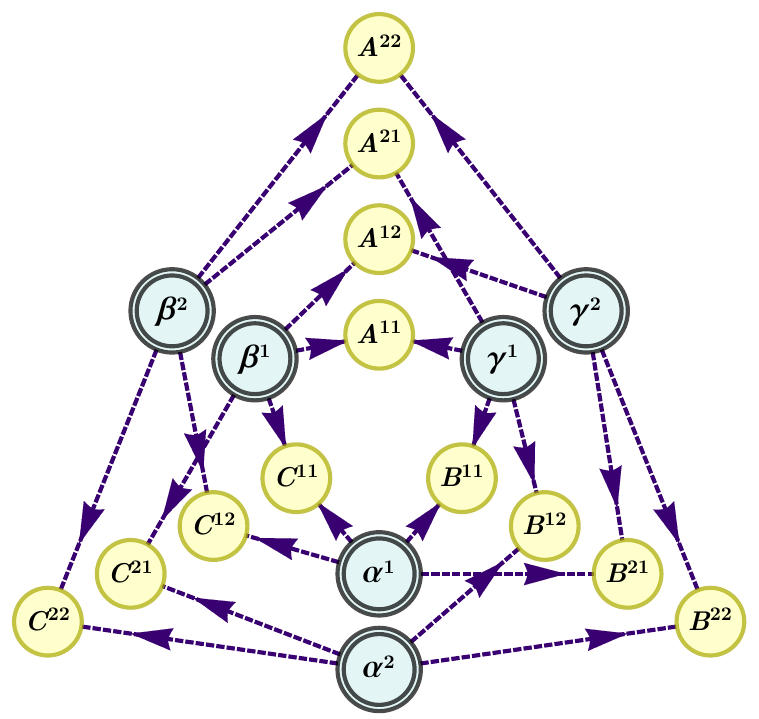}
	\hskip 2cm
	\includegraphics[width=0.3\columnwidth]{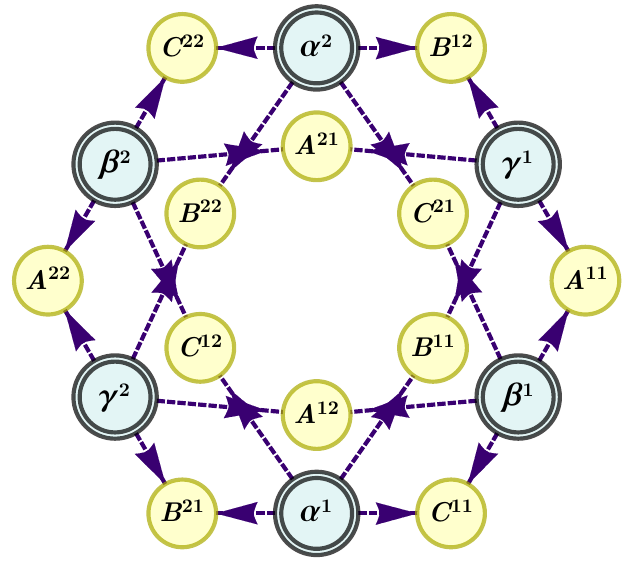}
	\caption{Two depictions of the second level in the inflation hierarchy of Ref.~\cite{navascues2017convergence} for the triangle scenario. It is also known as the web inflation in Refs.~\cite{krivachy2019neural,wolfe2019inflation}, or as the hexagon web inflation. In this inflation there are two copies of each source, and moreover each of these sources sends two copies of the shared variables, each one to a different copy of each of the parties. Therefore, there are four copies of each of the parties, each of which recieves one different pair of copies of the corresponding shared variables.}
	\label{fig:WebInflation}
\end{figure*}

Let us now use inflation arguments to derive necessary conditions for a distribution to admit a model of the type given by Eq.~\eqref{eq:trianglelocal} in the main text.
We begin assuming that such model exists, so we know the local variables $\alpha$, $\beta$ and $\gamma$, characterized by the distributions $\mu_{BC}(\alpha)$, $\mu_{AC}(\beta)$ and $\mu_{AB}(\gamma)$, respectively, and the responses $P_A(a|\beta,\gamma)$, $P_B(b|\alpha,\gamma)$ and $P_C(c|\alpha,\beta$) of Alice, Bob and Charlie, respectively, which all combined give rise to $P(a,b,c)$.
Having access to this information means that we can imagine duplicating the local variables and responses (by going to the manufacturer and buying new sources and measurement devices), and also cloning the information that the sources send to the parties.
Then, one could imagine arranging the available sources and parties in the inflated network depicted in Fig.~\ref{fig:WebInflation}.
The question is now, what are the properties of the distribution $p_\text{inf}(\{a^{i,j}\},\{b^{k,l}\},\{c^{m,n}\})$ that is generated in this new network?

For once, the distribution is well defined, i.e.,
\begin{align}
	p_\text{inf}(\{a^{i,j}\},\{b^{k,l}\},\{c^{m,n}\}) &\geq 0 \qquad\forall\,a^{i,j},b^{k,l},c^{m,n},i,j,k,l,m,n,\label{LPpos}\\
	\sum_{\{a^{i,j}\}}\sum_{\{b^{k,l}\}}\sum_{\{c^{m,n}\}}p_\text{inf}(\{a^{i,j}\},\{b^{k,l}\},\{c^{m,n}\}) &= 1.\label{LPnorm}
\end{align}

The assumption that the sources and parties are all indistinguishable copies implies that the distribution is invariant under a number of permutations of its elements.
As an illustration, performing the swap $\alpha^1\,{\leftrightarrow}\,\alpha^2$ in Fig.~\ref{fig:WebInflation}, which corresponds to the transformations $b^{k,1}\,{\leftrightarrow}\, b^{k,2}$ and $c^{1,n}\,{\leftrightarrow}\, c^{2,n}$, leaves the distribution invariant.
More generally, if $\pi$, $\pi'$ and $\pi''$ are three independent permutations, we have that $p_\text{inf}$ must satisfy
\begin{equation}
	p_\text{inf}(\{a^{\pi(i),\pi'(j)}\},\{b^{\pi'(k),\pi''(l)}\},\{c^{\pi''(m),\pi(n)}\}) - p_\text{inf}(\{a^{i,j}\},\{b^{k,l}\},\{c^{m,n}\}) = 0\qquad\forall\,a^{i,j},b^{k,l},c^{m,n},i,j,k,l,m,n,\pi,\pi',\pi''.
	\label{LPinf}
\end{equation}

Moreover, given that the sources and response functions in the inflation are copies of the elements in the original model assumed, there exist marginals of $p_\text{inf}$ that can be associated to elements of the original probability distribution $P(a,b,c)$.
Firstly, note that if one marginalizes over the copies that have different indices on the left and on the right, one is left with two copies of the original triangle scenario.
This means that
\begin{equation}
	\sum_{a^{1,2},a^{2,1}}\sum_{b^{1,2},b^{2,1}}\sum_{c^{1,2},c^{2,1}} p_\text{inf}(\{a^{i,j}\},\{b^{k,l}\},\{c^{m,n}\}) = P(a^{1,1},b^{1,1},c^{1,1})\,P(a^{2,2},b^{2,2},c^{2,2})\qquad\forall\,a^{1,1},a^{2,2},b^{1,1},b^{2,2},c^{1,1},c^{2,2}.
	\label{LPhierarchy}
\end{equation}
\begin{figure*}[t!]
	\hfill
	\subfloat[\label{fig:hierarchy}]{
		\centering
    \begin{minipage}[t]{0.32\textwidth}
    	\centering
			\includegraphics[scale=0.15,trim={0cm 0.1cm 0cm 0cm},clip]{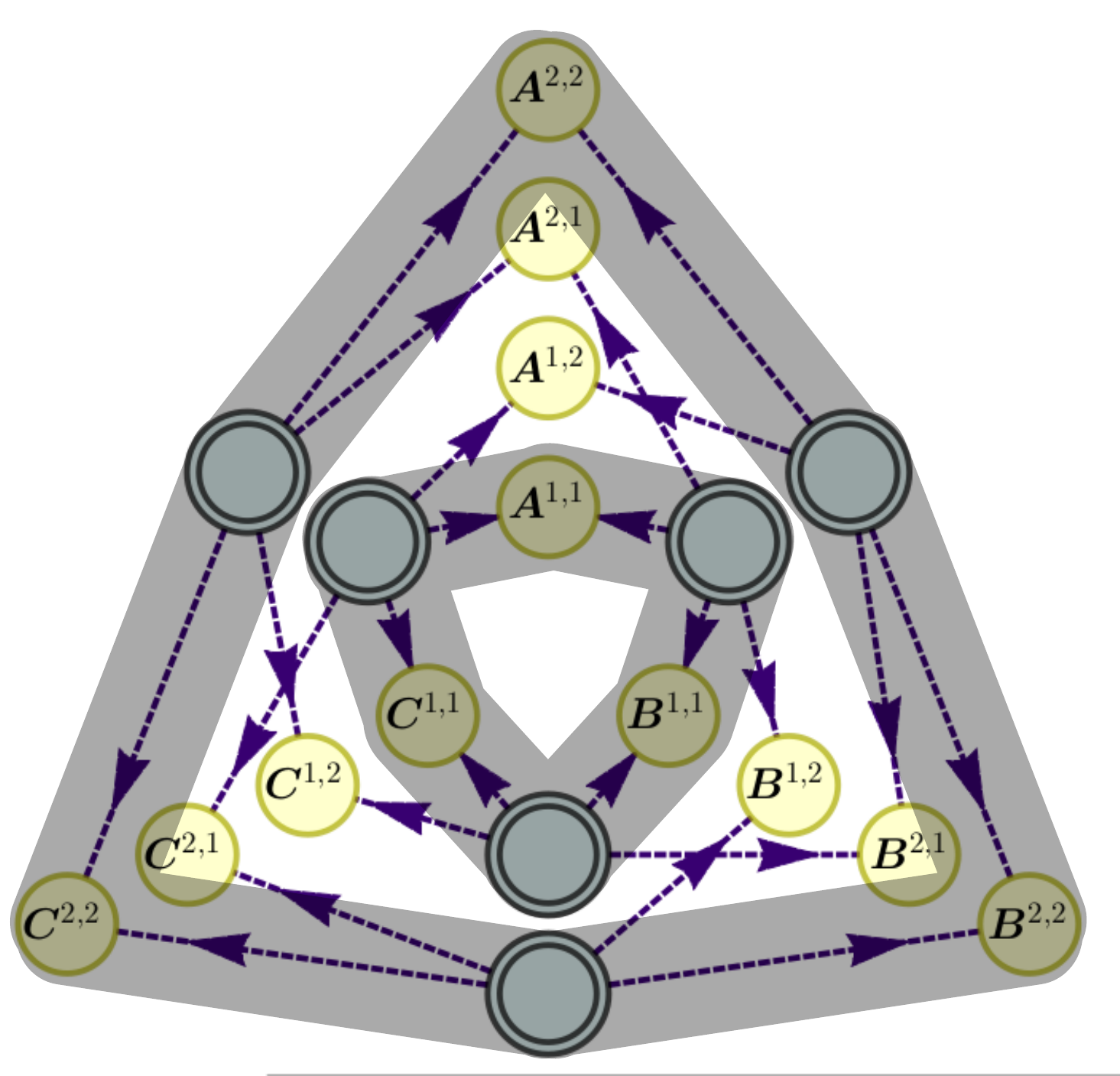}\\
			$p_\text{inf}(a^{1,1},a^{2,2},b^{1,1},b^{2,2},c^{1,1},c^{2,2})=P(a^{1,1},b^{1,1},c^{1,1})P(a^{2,2},b^{2,2},c^{2,2})$
 		\end{minipage}
	}
	\hfill
	\subfloat[\label{fig:higherorder}]{
		\centering
    \begin{minipage}[t]{0.32\textwidth}
    	\centering
			\includegraphics[scale=0.15,trim={0cm 0.1cm 0cm 0cm},clip]{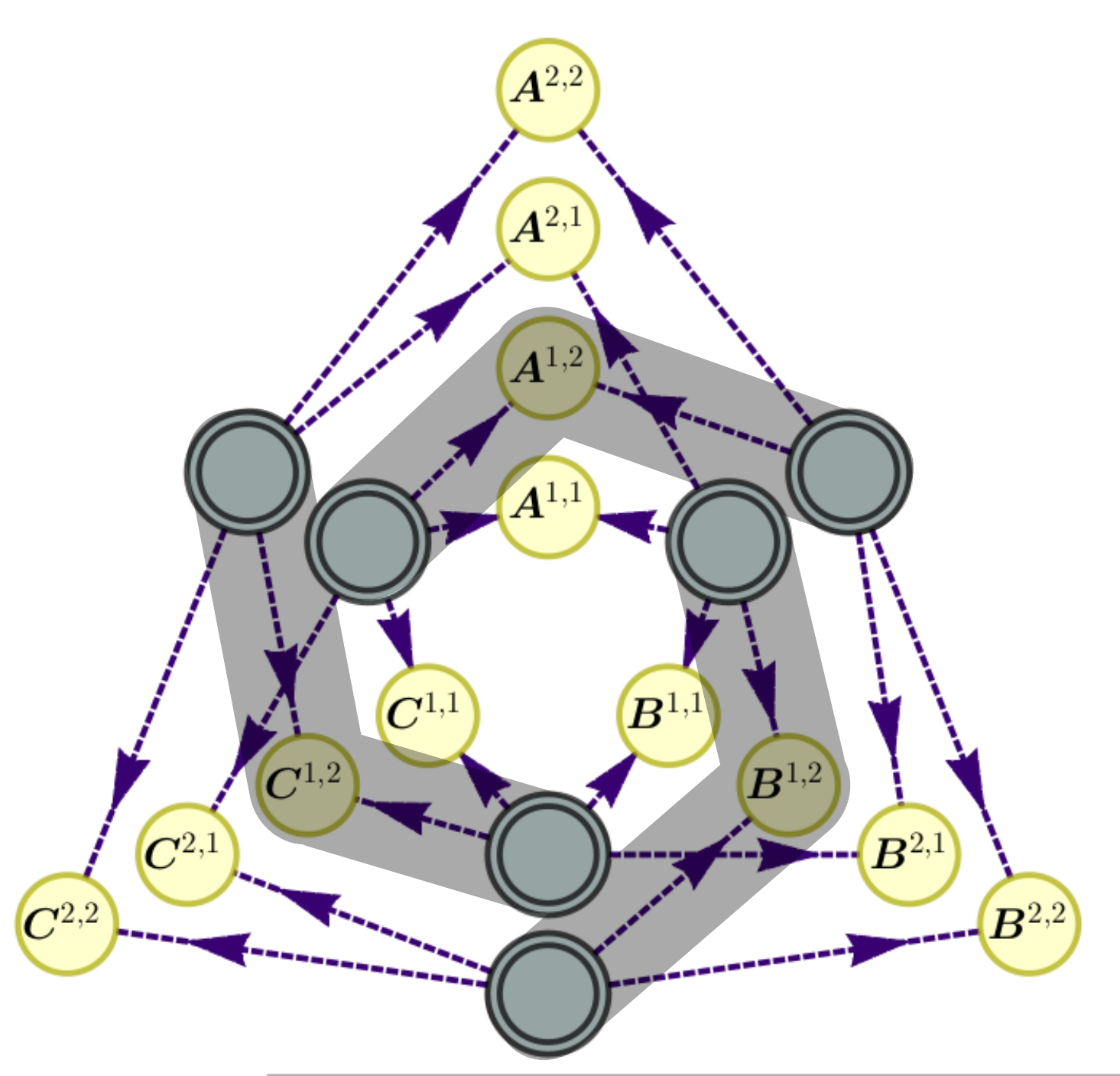}\\
			$p_\text{inf}(a^{1,2},b^{1,2},c^{1,2})=P(a^{1,2})P(b^{1,2})P(c^{1,2})$
		\end{minipage}
	}
	\hfill
	\subfloat[\label{fig:lpi}]{
		\centering
    \begin{minipage}[t]{0.32\textwidth}
    	\centering
			\includegraphics[scale=0.15,trim={0cm 0.1cm 0cm 0cm},clip]{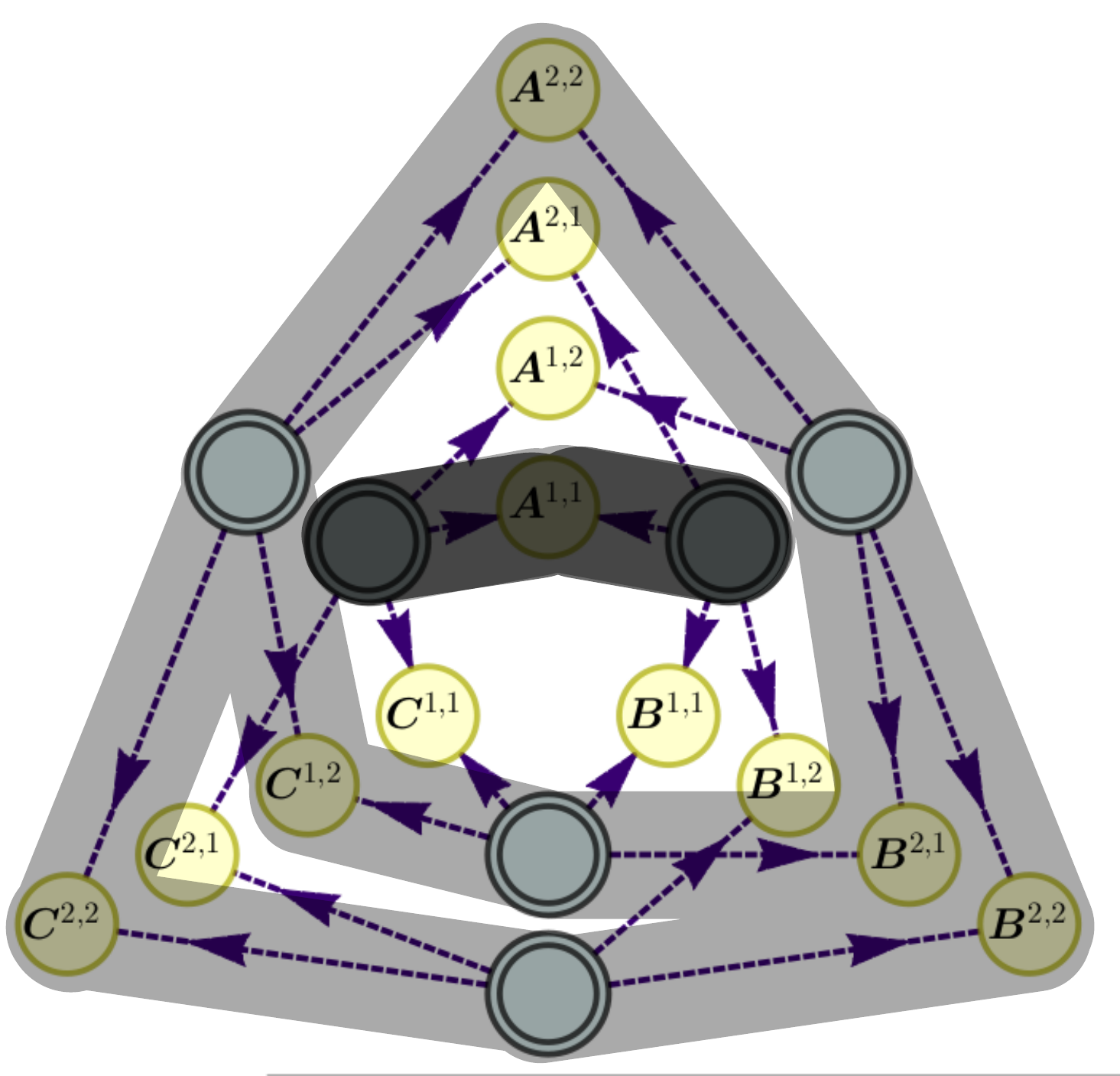}\\
			$p_\text{inf}(a^{1,1},a^{2,2},b^{2,1},b^{2,2},c^{1,2},c^{2,2})=P(a^{2,2})p_\text{inf}(a^{1,1},b^{2,1},b^{2,2},c^{1,2},c^{2,2})$
		\end{minipage}
	}
	\hfill
	\caption{Types of marginals of the probability distribution in the inflation, $p_\text{inf}$, that can be associated to the original distribution. \protect\subref{fig:hierarchy} represents the marginals used in the hierarchy constraints of Eq.~\eqref{LPhierarchy}. These are sufficient to describe a convergent inflation hierarchy, but the remaining provide tighter characterizations at a fixed inflation level. \protect\subref{fig:higherorder} represents the marginals used in the higher-order relations of Eq.~\eqref{LPhigherorder}. \protect\subref{fig:lpi} represents the marginals used in the linearized polynomial identification constraints of Eq.~\eqref{LPlpi}. Note that, in the latter, the whole dashed marginal is factorized in two: the stronger one, which is substituted by a known marginal of the original probability distribution, and the lighter one, which is a marginal of the inflation distribution by itself.}
	\label{fig:inflationConstraints}
\end{figure*}
Similarly, another constraint satisfied is
\begin{equation}
	\sum_{a^{1,1},a^{2,1},a^{2,2}}\sum_{b^{1,1},b^{2,1},b^{2,2}}\sum_{c^{1,1},c^{2,1},c^{2,2}} p_\text{inf}(\{a^{i,j}\},\{b^{k,l}\},\{c^{m,n}\}) = P(a^{1,2})\,P(b^{1,2})\,P(c^{1,2})\qquad\forall\,a^{1,2},b^{1,2},c^{1,2},
	\label{LPhigherorder}
\end{equation}
where the right-hand side represents a product of single-party marginals of the original distribution $P(a,b,c)$.
Note that Eqs.~(\ref{LPhierarchy}-\ref{LPhigherorder}) relate combinations of probabilities in the inflation to products of the elements of the original probability distribution.
It is possible also to consider more convoluted relations, where the right-hand side contains elements of the distribution in the inflation as well.
Namely, one can consider the constraints
\begin{equation}
	\sum_{\substack{a^{1,2}\\a^{2,1}}}\sum_{\substack{b^{1,1}\\b^{1,2}}}\sum_{\substack{c^{1,1}\\c^{2,1}}} p_\text{inf}(\{a^{i,j}\},\{b^{k,l}\},\{c^{m,n}\}) - P(a^{1,1})\sum_{\substack{a^{1,1}\\a^{1,2}\\a^{2,1}}}\sum_{\substack{b^{1,1}\\b^{1,2}}}\sum_{\substack{c^{1,1}\\c^{2,1}}} p_\text{inf}(\{a^{i,j}\},\{b^{k,l}\},\{c^{m,n}\})=0\quad\forall\,a^{1,1},a^{2,2},b^{2,1},b^{2,2},c^{1,2},c^{2,2},
	\label{LPlpi}
\end{equation}
plus the equivalent under cyclic permutation of parties.

The constraints in Eq.~\eqref{LPhierarchy} are sufficient to define a hierarchy of inflations that asymptotically converges to the characterization of all $P(a,b,c)$ compatible with Eq.~\eqref{eq:trianglelocal} in the main text \cite{navascues2017convergence}.
However, the additional constraints \eqref{LPhigherorder} and \eqref{LPlpi} have proved useful for further constraining the characterizations offered by inflation under restricted computational resources.
The constraints in Eq.~\eqref{LPhigherorder}, denoted as higher-degree relations in Ref.~\cite{navascues2017convergence}, allow to identify the W distribution as incompatible with Eq.~\eqref{eq:trianglelocal} in the main text \cite{wolfe2019inflation}.
In turn, Eq.~\eqref{LPlpi} are the linearized polynomial identification constraints.
These constraints have been used in Ref.~\cite{Gisin2020} for providing tighter characterizations of distributions compatible with the triangle scenario, and a comparison between the characterizations obtained with and without these constraints can be found in Ref.~\cite{alexThesis}.

\begin{figure}[h!]
	\hfill
	\subfloat[\label{fig:region}]{
		\centering
    \begin{minipage}[t]{0.23\textwidth}
    	\centering
			\includegraphics[width=0.9\textwidth]{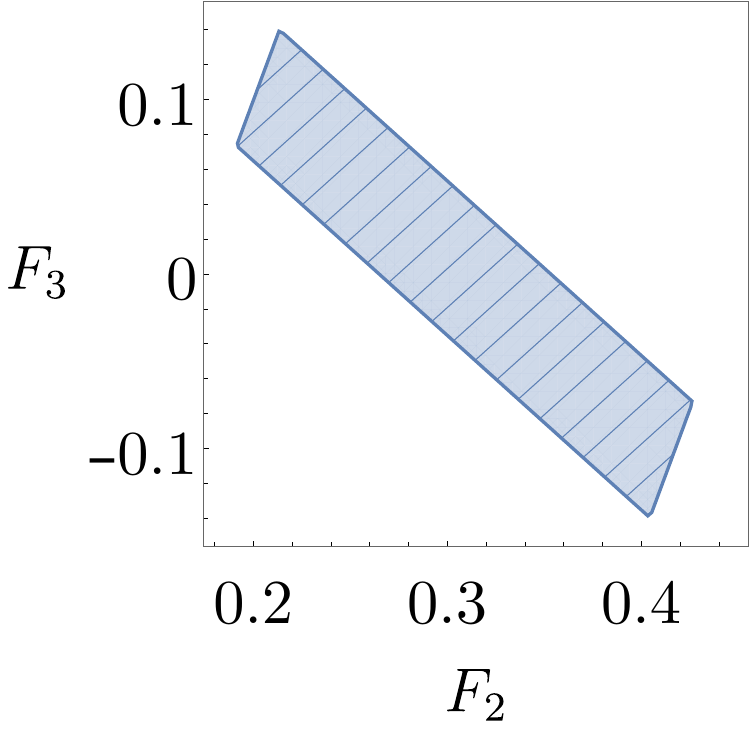}
 		\end{minipage}
	}
	\hfill
	\subfloat[\label{fig:regionWithHierarchy}]{
		\centering
    \begin{minipage}[t]{0.23\textwidth}
    	\centering
			\includegraphics[width=0.9\textwidth]{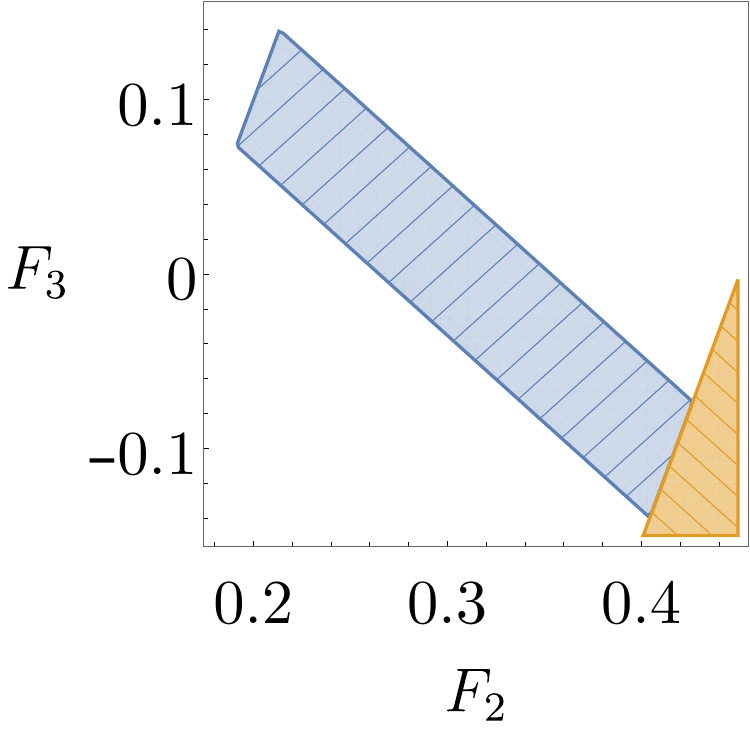}
		\end{minipage}
	}
	\hfill
	\subfloat[\label{fig:regionWithHigherOrder}]{
		\centering
    \begin{minipage}[t]{0.23\textwidth}
    	\centering
			\includegraphics[width=0.9\textwidth]{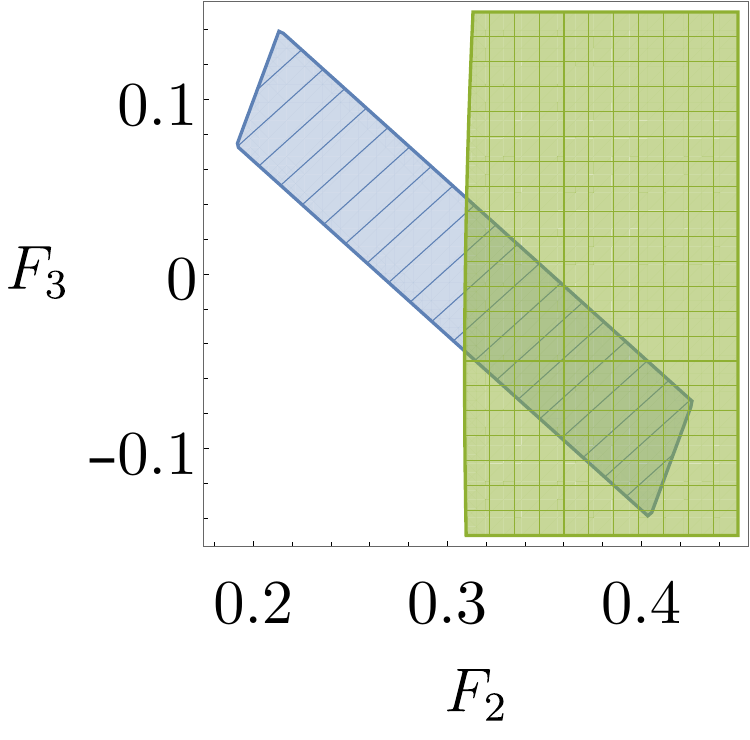}
		\end{minipage}
	}
	\hfill
	\subfloat[\label{fig:regionWithLPI}]{
		\centering
    \begin{minipage}[t]{0.23\textwidth}
    	\centering
			\includegraphics[width=0.9\textwidth]{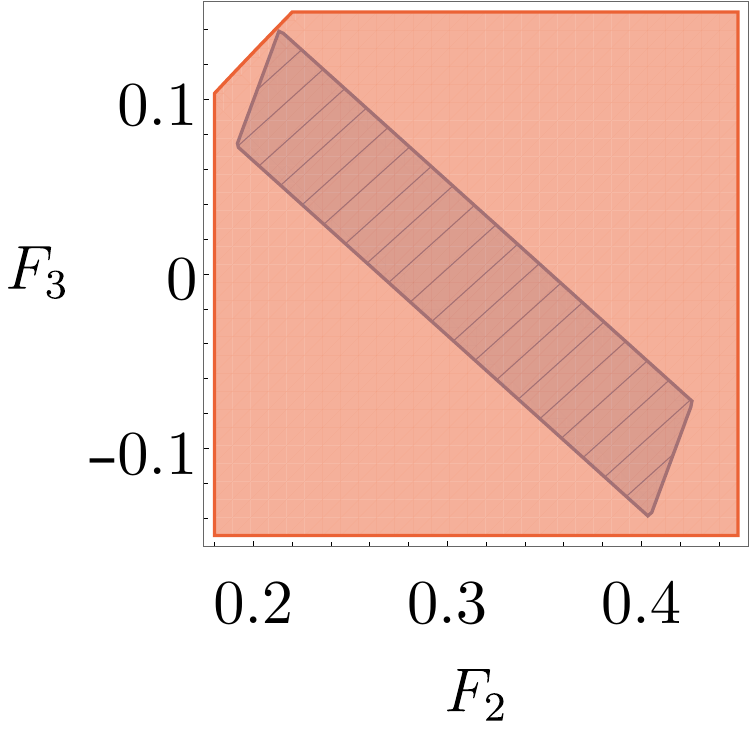}
		\end{minipage}
	}
	\hfill
	\caption{Example of regions of the projection of the feasible polytope, $\mathcal{P}_u$, in the region defined by $(F_{AB}\,{=}\,F_{BC}\,{=}\,F_{AC}\,{\equiv}\, F_2$, $F_{ABC}\equiv F_3)$, for which the family of distributions $q_u^{-}(i,j,k)$ is identified not to admit a triangle-local model of the form of Eq.~\eqref{eq:trianglelocal} in the main text by the inflation linear programs with different types of constraints.
	\protect\subref{fig:region} depicts $\mathcal{P}_{0.8090}$ in blue.
	The orange region in \protect\subref{fig:regionWithHierarchy} represents the distributions proved not to admit a model based on the standard inflation hierarchy, given by Eqs.~(\ref{LPpos}-\ref{LPhierarchy}).
	The green region in \protect\subref{fig:regionWithHigherOrder} represents the triangle-nonlocal distributions identified when the higher-order constraints of Eq.~\eqref{LPhigherorder} are added to the former.
	The red region in \protect\subref{fig:regionWithLPI} represents the distributions that are proved not to admit a triangle-local model after adding the LPI constraints of Eq.~\eqref{LPlpi}.
	Note that the linear program \eqref{LP} is infeasible for every point inside $\mathcal{P}_{0.8090}$.
	Following the arguments in the main text, this is a proof that the distribution in Eq.~\eqref{eq:genuine} in the main text is triangle nonlocal for $u\,{=}\,0.8090$.}
	\label{fig:constraintCertificates}
\end{figure}

Recall that the conditions (\ref{LPpos}-\ref{LPlpi}) are consequences of the correlations admitting a model of the type of Eq.~\eqref{eq:trianglelocal} in the main text.
Thus, in order to see whether a distribution $P(a,b,c)$ admits such a model, one can consider the problem
\begin{equation}
	\text{find } p_\text{inf} \text{ such that } \eqref{LPpos},\eqref{LPnorm},\eqref{LPinf},\eqref{LPhierarchy},\eqref{LPhigherorder},\eqref{LPlpi},
	\label{LP}
\end{equation}
which, since all constraints are linear once $P(a,b,c)$ is defined, can be formulated as a linear program.
If a solution to \eqref{LP} does not exist, then it is certified that the premise, namely that $P(a,b,c)$ satisfies Eq.~\eqref{eq:trianglelocal} in the main text, is false.
The absence of a solution for Eq.~\eqref{LP} can be demonstrated, via Farka's lemma, in terms of a certificate of infeasibility.
In Fig.~\ref{fig:constraintCertificates} we show an exemplification of the certificates obtained when the different types of constraints are added to the linear program.

\section{Noise-robust witnesses of triangle nonlocality}\label{app:continuum}
In order to extend our derivation to the continuum, and thus prove that $P_u(a,b,c)$ is triangle-nonlocal in the whole range $0.7504\,{\leq}\,u\,{\leq}\,0.8101$, let us state explicitly Farkas' lemma~\cite{FarkasLemma}:
\begin{equation}
	\not\exists\,\bm{x}\text{ s.t. }A\cdot\bm{x}\geq\bm{b}\Leftrightarrow\exists\,\bm{y}\text{ s.t. }\left\{\begin{split}\bm{y}\geq &\,\bm{0},\\ \bm{y}\cdot A=&\,\bm{0},\\ \bm{y}\cdot\bm{b} > &\,0.\end{split}\right.
	\label{eq:farkas}
\end{equation}
This is, if a linear program $[\text{find }\bm{x}\text{ s.t. }A\cdot\bm{x}\geq\bm{b}]$ does not admit any feasible solution $\bm{x}$, Farkas' lemma guarantees the existence of a vector $\bm{y}\,{\geq}\,\bm{0}$ that satisfies $\bm{y}\cdot A\,{=}\,\bm{0}$ and $\bm{y}\cdot\bm{b} \,{>}\, 0$.
In the programs we consider [see Eq.~\eqref{LP} in Appendix~\ref{app:inflation}], the vector $\bm{x}\equiv p_\text{inf}$ contains the probability elements of the distribution in the inflation, the vector $\bm{b}$ is built from the right-hand sides of Equations \eqref{LPpos}-\eqref{LPlpi}, and the matrix $A$ contains the coefficients accompanying the probability elements in the left-hand sides of Equations \eqref{LPpos}-\eqref{LPlpi}.
Note that all the elements in $A$ are independent of the distribution under scrutiny, except for those encoding the LPI constraints.
For these, some coefficients in the corresponding rows of $A$ are (particularizing to the distributions $q_u^t$) $c^t_u$ or $1-c^t_u$, where $c^t_u\,{=}\,\sum_{j,k}q^t_u(1,j,k)\,{=}\,\sum_{i,j}q^t_u(i,j,1)\,{=}\,\sum_{i,k}q^t_u(i,1,k)\,{=}\,(1-t+2u^2)/2$.
This is the reason behind the argument, presented in the section \textit{Proof for discrete values $u<u_0$} in the main text, that the witnesses obtained are only applicable to a fixed $u$ but to any allowed values of $F_{AB}$, $F_{AC}$, $F_{BC}$ and $F_{ABC}$, since $A=A(u)$ but the condition $\bm{y}_u\cdot A(u)\,{=}\,\bm{0}$ is not robust to perturbations in $u$.

However, one does not need such strict conditions in order to have a guarantee of infeasibility.
Note that, for vectors $\bm{x}$ satisfying $\bm{x}\geq\bm{0}$ and $\sum_i x_i=1$ (such as those considered in this work, since $\bm{x}\equiv p_\text{inf}$ represents a probability distribution), the two last properties of $\bm{y}$ in Eq.~\eqref{eq:farkas} can be combined into $\bm{y}\cdot\bm{b}>\max(\bm{y}\cdot A)$.
This expression, in contrast with the properties in Eq.~\eqref{eq:farkas}, is robust to perturbations in the coefficients of both $A$ and $\bm{b}$.
Yet, note that if for some $\bm{y}$ it is satisfied that $\bm{y}\cdot\bm{b}>\max(\bm{y}\cdot A)$, then no $\bm{x}$ (positive and normalized) exists that satisfies $A\cdot\bm{x}\geq \bm{b}$: multiplying this expression by $\bm{y}$ produces a contradiction with $\bm{y}\cdot\bm{b}>\max(\bm{y}\cdot A)$.
This means that, when addressing the compatibility of a tripartite probability distribution with triangle-local models via Eq.~\eqref{LP}, if a certificate of infeasibility $\bm{y}$ is found for that particular distribution (which proves that the distribution in question is triangle nonlocal), satisfying the inequality $\bm{y}\cdot\bm{b}'>\max(\bm{y}\cdot A')$ is a witness of the triangle nonlocality of the distribution used to build $\bm{b}'$ and $A'$.
In other words, since Eqs.~(\ref{LPpos}-\ref{LPlpi}) are necessary conditions for a tripartite distribution $P$ to have a triangle-local model, when the components of $A$ and $\bm{b}$ are written in terms of a generic probability distribution [$P$ in Eqs.~(\ref{LPhierarchy}-\ref{LPlpi})], the quantity $\bm{y}\cdot\bm{b}\leq\max(\bm{y}\cdot A)$ is a Bell-like polynomial inequality whose violation certifies triangle nonlocality.

Now, as discussed above, in the concrete problem of determining the triangle nonlocality of $q_u^t(i,j,k)$, the coefficients of $A$ only depend on $u$ and are continuous functions of $u$ for fixed $t$.
This is also the case for the coefficients of $\bm{b}$ when particularized to the vertices of $\mathcal{P}_u$.
In general, the coefficients of $\bm{b}$ are continuous functions of $u$, $F_{AB}$, $F_{AC}$, $F_{BC}$ and $F_{ABC}$, but this has no impact in the arguments below.
Because of the continuity in $u$, after finding a certificate $\bm{y}_u$ that satisfies $\bm{y}_u\cdot A(u)=\bm{0}$ and $\bm{y}_u\cdot\bm{b}(u) > 0$ for a fixed value of $u$, the quantity $\bm{y}_u\cdot\bm{b}(u')-\max[\bm{y}_u\cdot A(u')]$ is a continuous function in $u'$ that witnesses triangle nonlocality wherever it takes positive values.
The fact that the quantity $\bm{y}_u\cdot\bm{b}(u')-\max[\bm{y}_u\cdot A(u')]$ is a continuous function of $u'$ guarantees, in particular, that if $P_u(a,b,c)$ is triangle-nonlocal for some value of $u$, there exists a neighborhood around $u$ for which the corresponding $P_{u'}(a,b,c)$ is triangle-nonlocal as well.
The procedure that we implement in the computational appendix~\cite{compAppendix} is based on this observation.
Recursively, the program proves the triangle nonlocality of $P_u(a,b,c)$ for a particular value of $u$ by finding a Farkas' certificate $\bm{y}_u$, computes the furthest value $u'$ for which the condition $\max\left[\bm{y}_u\cdot A(u')\right]\,{<}\,\bm{y}_u\cdot\bm{b}(u')$ guarantees that $P_{u'}(a,b,c)$ is triangle-nonlocal as well, and sets $u\,{=}\,u'$ to begin again.
However, as noted above, the quantity $\bm{y}_u\cdot\bm{b}\leq\max(\bm{y}_u\cdot A)$ for $A$ and $\bm{b}$ described by the left-hand and right-hand sides of Eqs.~(\ref{LPpos}-\ref{LPlpi}), respectively, are general witnesses of triangle nonlocality for binary-outcome probability distributions.
All the witnesses of triangle nonlocality for arbitrary distributions extracted in the process can be found in the computational appendix~\cite{compAppendix} as well.

\end{document}